\documentstyle[11pt,amssymb,epsf]{article}
\makeatletter
\@addtoreset{equation}{section}
\makeatother

\def\crampest{\medmuskip = 1mu plus 1mu minus 1mu}

\def\ben{\begin{equation}}
\def\een{\end{equation}}

  \let\n=\nu  \let\p=\pi

\let\C=\Chi

\def\nn{\nonumber} \def\bd{\begin{document}} \def\ed{\end{document}}
\def\ds{\documentstyle} \let\fr=\frac \let\bl=\bigl \let\br=\bigr
\let\Br=\Bigr \let\Bl=\Bigl
\let\bm=\bibitem
\let\na=\nabla
\let\pa=\partial \let\ov=\overline
\newcommand{\be}{\begin{equation}}
\newcommand{\ee}{\end{equation}}
\def\ba{\begin{array}}
\def\ea{\end{array}}
\def\ft#1#2{{\textstyle{{\scriptstyle #1}\over {\scriptstyle #2}}}}
\def\fft#1#2{{#1 \over #2}}
\def\del{\partial}
\def\vp{\varphi}
\def\sst#1{{\scriptscriptstyle #1}}
\def\oneone{\rlap 1\mkern4mu{\rm l}}
\def\td{\tilde}
\def\wtd{\widetilde}
\def\ie{{\it i.e.\ }}
\def\dalemb#1#2{{\vbox{\hrule height .#2pt
        \hbox{\vrule width.#2pt height#1pt \kern#1pt
                \vrule width.#2pt}
        \hrule height.#2pt}}}
\def\square{\mathord{\dalemb{6.8}{7}\hbox{\hskip1pt}}}
\newcommand{\ho}[1]{$\, ^{#1}$}
\newcommand{\hoch}[1]{$\, ^{#1}$}
\newcommand{\bea}{\begin{eqnarray}}
\newcommand{\eea}{\end{eqnarray}}
\newcommand{\ra}{\rightarrow}
\newcommand{\lra}{\longrightarrow}
\newcommand{\Lra}{\Leftrightarrow}
\newcommand{\bp}{\tilde \beta^\prime}
\newcommand{\tr}{{\rm tr} }
\newcommand{\Tr}{{\rm Tr} }
\def\0{{\sst{(0)}}}
\def\1{{\sst{(1)}}}
\def\2{{\sst{(2)}}}
\def\3{{\sst{(3)}}}
\def\4{{\sst{(4)}}}
\def\5{{\sst{(5)}}}
\def\6{{\sst{(6)}}}
\def\7{{\sst{(7)}}}
\def\8{{\sst{(8)}}}
\def\n{{\sst{(n)}}}
\def\cA{{{\cal A}}}
\def\cB{{{\cal B}}}
\def\cF{{{\cal F}}}
\def\cH{{{\cal H}}}
\def\tV{\widetilde V}
\def\tW{\widetilde W}
\def\tH{\widetilde H}
\def\tE{\widetilde E}
\def\tF{\widetilde F}
\def\tA{\widetilde A}
\def\im{{{\rm i}}}
\def\tY{{{\wtd Y}}}
\def\ep{{\epsilon}}
\def\vep{{\varepsilon}}
\def\R{\rlap{\rm I}\mkern3mu{\rm R}}
\def\bD{{{\bar D}}}

\def\R{\rlap{\rm I}\mkern3mu{\rm R}}
\def\bD{{{\bar D}}}
\def\R{{{\Bbb R}}}
\def\C{{{\Bbb C}}}
\def\H{{{\Bbb H}}}
\def\CP{{{\Bbb C}{\Bbb P}}}
\def\RP{{{\Bbb R}{\Bbb P}}}
\def\Z{{{\Bbb Z}}}
\def\bA{{{\Bbb A}}}
\def\bB{{{\Bbb B}}}
\def\bC{{{\Bbb C}}}
\def\bD{{{\Bbb D}}}
\def\bE{{{\Bbb E}}}
\def\bZ{{{\Bbb Z}}}
\def\Re{{{\frak{Re}}}}
\def\Im{{{\frak{Im}}}}
\def\cosec{{\,\hbox{cosec}\,}}
\def\Gm{{\Gamma_{\!\! -}}}
\def\Gp{{\Gamma_{\!\! +}}}
\def\stan{{standard }}
\def\nonstan{{supernumerary }}
\def\p{{\partial}}

\def\bog{{Bogomolny }}

\newcommand{\tamphys}{\it Center for Theoretical Physics,
Texas A\&M University, College Station, TX 77843}

\newcommand{\upenn}{\it Department of Physics and Astronomy,\\ University
of Pennsylvania, Philadelphia, PA 19104}

\newcommand{\brussels}{\it Physique Th\'eorique et Math\'ematique,
Universit\'e Libre de Bruxelles,\\ Campus Plaine C.P. 231, B-1050
Bruxelles, Belgium}

\newcommand{\damtp}{\it DAMTP, Centre for Mathematical Sciences,\\
 Cambridge University, Wilberforce Road, Cambridge CB3 OWA, UK}

\newcommand{\auth}{M. Cveti\v c\hoch{*2}, G.W. Gibbons\hoch{\dagger},   
H. L\"u\hoch{\ddagger1} and C.N. Pope\hoch{\ddagger1}}

\begin{document}

\begin{flushright}
DAMTP-2005-39\ \ \ MIFP-05-08\ \ \ UPR-1114-T \\
{\bf hep-th/0504080}\\
April\  2005
\end{flushright}

%\vspace{10pt}

\begin{center}

{\large {\bf Rotating Black Holes in Gauged Supergravities;
Thermodynamics, Supersymmetric Limits, Topological Solitons and Time Machines}}

\vspace{20pt}
\auth

\vspace{10pt}{\hoch{*}\it Department of Physics and Astronomy,\\
University of Pennsylvania, Philadelphia, PA 19104, USA}

\hoch{\dagger}\damtp

\vspace{10pt}{\hoch{\ddagger}\it George P. \& Cynthia W. Mitchell
Institute for Fundamental Physics,\\ Texas A\& M University,
College Station, TX 77843-4242, USA}

%\vspace{10pt} {\hoch{\ddagger}\it Interdisciplinary Center for
%Theoretical Study, \\ University of Science \& Technology of China,
%Hefei, Anhui 230026, China}

%{\it \hoch{*} Michigan Center for Theoretical Physics\\
%University of Michigan, Ann Arbor, MI 48109-1120}
%
%
%
%\vspace{10pt} {\hoch{\dagger}\brussels}

\vspace{20pt}

\underline{ABSTRACT}
\end{center}

%\begin{abstract}

    We study the thermodynamics of the recently-discovered
non-extremal charged rotating black holes of gauged supergravities in
five, seven and four dimensions, obtaining energies, angular momenta
and charges that are consistent with the first law of thermodynamics.
We obtain their supersymmetric limits by using these expressions
together with an analysis of the AdS superalgebras including
R-charges.  We give a general discussion of the global structure of
such solutions, and apply it in the various cases.  We obtain new
regular supersymmetric black holes in seven and four dimensions, as
well as reproducing known examples in five and four dimensions.  We
also obtain new supersymmetric non-singular topological solitons in
five and seven dimensions. The rest of the supersymmetric solutions
either have naked singularities or naked time machines. The latter can
be rendered non-singular if the asymptotic time is periodic.  This
leads to a new type of quantum consistency condition, which we call a
{\it Josephson quantisation condition}. Finally, we discuss some
aspects of rotating black holes in G\"odel universe backgrounds.

%\end{abstract}

{\vfill\leftline{}\vfill \vskip 10pt \footnoterule {\footnotesize
{\footnotesize
\hoch{1} Research supported in part by DOE grant
DE-FG03-95ER40917 and NSF grant INTO3-24081.}\vskip 2pt
\hoch{2} Research supported in part by DOE grant
DE-FG02-95ER40893, NSF grant INTO3-24081, and the\\
$\phantom{xxxxi}$  Fay R. and Eugene L.
Langberg Chair.}\vskip 2pt
}

\pagebreak

\tableofcontents
\addtocontents{toc}{\protect\setcounter{tocdepth}{2}}
\newpage

\section{Introduction}

   With the development of the AdS/CFT correspondence
\cite{mal,guklpo}, it has become important to investigate backgrounds
in gauged supergravities describing charged black holes.  Of
particular interest are five-dimensional gauged supergravity, in the
context of the type IIB string, and the gauged supergravities in four
dimensions and seven dimensions, in the context of M-theory.
Supersymmetric black holes in the AdS background play a particularly
important role.  Typically, it turns out that these are singular
unless they are in addition rotating, and so it becomes essential to
study rotating charged black holes in gauged supergravity.

    A point of special interest is the occurrence of closed timelike
curves in supersymmetric solutions, including solutions with a large
amount of supersymmetry. This raises the issue of whether Chronology
Protection may be achieved by some stringy or quantum mechanical
consistency condition.  For example is it associated with a breakdown
of unitarity of the boundary field theory?  It is well known that the
Kerr solution has closed timelike curves and it remains an open
question whether they play a role in realistic models of gravitational
collapse.  Supergravity solutions with closed timelike curves offer
the opportunity of studying these questions within a well defined and
controlled theoretical context.

   The general solutions for non-extremal charged rotating black holes
in ungauged supergravities have been obtained, in five dimensions
\cite{cvetyoum}, four dimensions \cite{cy1}, and then in other
dimensions \cite{cy2,cvetyoum2}. The thermodynamics and grey body
factors of general five and four dimensional solutions we studied in
\cite{cl1} and \cite{cl2}, respectively. The global properties of a
special case in five dimensions \cite{BMPV}, where the three electric
charges of the general solution are set equal, have been studied in
considerable detail \cite{gibher}.  They can in fact be constructed by
a rather mechanical solution-generating procedure, involving the use
of global symmetries of the ungauged supergravity theories.  In this
construction, one begins with the standard uncharged rotating black
hole solutions, and introduces the charges by making global symmetry
transformations.  The extremal limits of these charged rotating black
holes provide supersymmetric BPS configurations.

   Constructing the analogous charged rotating solutions in gauged
supergravity is a much more challenging problem.  The general
uncharged AdS black hole solutions were found in four dimensions in
\cite{carter1}, in five dimensions in \cite{hawhuntay}, and in
all dimensions $D\ge 6$ in \cite{gilupapo1,gilupapo2}.  However, in
the gauged supergravities one no longer has global symmetries that can
be used in order to introduce charges.  Instead, there is little
option but to resort to more ``brute force'' methods for solving the
supergravity equations.  By such methods, various non-extremal charged
rotating black hole solutions have been constructed in gauged
supergravities in four, five and seven dimensions.  The simplest case,
obtained long ago in \cite{carter3}, is the Kerr-Newman-AdS
black hole in four dimensions, which as a solution of the
Einstein-Maxwell system with a cosmological constant, can be viewed
also as a solution in four-dimensional gauged ${\cal N}=2$
supergravity.

   Recently, non-extremal charged rotating black hole solutions in
five-dimensional gauged supergravity have been constructed.  In all of
these, the problem was simplified by taking the two rotation
parameters of a general five-dimensional rotating black hole to be
equal.  First, the solution in five-dimensional gauged ${\cal N}=2$
(minimal) supergravity was found \cite{d5gauge1}.  This was
generalised to the case of maximal gauged supergravity, with the black
holes carrying three independent electric charges, in
\cite{d5gauge2}.  Setting the three charges equal reduces to the case
studied previously in \cite{d5gauge1}.

   Generalisations of the Kerr-Newman-AdS black holes in four dimensions
were then found \cite{d4gauge}, which can be viewed as solutions in
gauged ${\cal N}=4$ supergravity, with two independent electric (or magnetic)
charges.  If these are set equal, the solutions reduce to the Kerr-Newman-AdS
black holes.  In terms of gauged ${\cal N}=8$ supergravity, the solutions
in \cite{d4gauge} correspond to 4-charge black holes, in which the
charges are set pairwise equal.

   In seven dimensions, non-extremal charged rotating black hole
solutions in the maximal gauged ${\cal N}=4$ supergravity were obtained
in \cite{d7gauge}, with two independent electric charge parameters.  Again,
the problem was simplified by taking the three rotation parameters of
a general seven-dimensional black hole to be equal.

   Certain supersymmetric rotating black holes in gauged supergravities have
also been previously obtained.  Specifically, in four dimensions it
was shown in \cite{kosper} that one can take a BPS limit of the
Kerr-Newman-AdS black hole.  In fact in general this BPS limit is not a
black hole at all, since it lacks an event horizon and thus it has a
naked singularity.  However, as shown in \cite{kosper}, if a certain
relation between the angular momentum and the charge holds, one obtains
a genuine supersymmetric rotating black hole. 

   In five dimensions, supersymmetric solutions including black holes 
were obtained in gauged supergravity by a direct approach, in which
the supersymmetry was imposed from the outset.  Examples included
solutions found by Klemm and Sabra \cite{klemm1,klemm2}, which contain
naked singularities or closed timelike curves, and solutions found
by Gutowski and Reall \cite{gutrea1,gutrea2}, which have regular
horizons.

   The plan of the paper is as follows.  We begin in section 2 by
recalling some basic definitions associated with horizons and closed
timelike curves (CTC's), using the BMPV black hole \cite{BMPV} as an
illustrative example.  We pay special attention to the asymptotically
AdS case and the properties of Killing vectors arising from Killing
spinors.  We show how the metrics we are considering allow a complete
integration of the geodesics using Hamilton-Jacobi theory, and this
allows us to give a simple physical derivation of the second law of
thermodynamics for black holes. At the end of section 2, we give a
derivation of a new type of quantisation condition that arises when
ensuring that solutions are globally well-defined.  This new
condition arises only if the time coordinate is
periodic.  Because an analogous condition arises when considering the
time-dependent Josephson effect, we propose calling this new
quantisation condition the Josephson quantisation condition. (See
\cite{weinberg} for a related discussion of the standard
time-dependent Josephson effect using just the general ideas of
gauge-invariance rather than a particular microscopic model as
in Josephson's original derivation.)

   In the following three sections, we study rotating black holes in
$D=5$, $D=7$ and $D=4$ gauged supergravities, and their supersymmetric
limits.  We find many interesting such limits, but the most important
outcomes are new regular supersymmetric rotating black holes in seven
dimensions and in four dimensions, and regular topological solitons in
$D=5$ and $D=7$.  In detail, these sections are as follows.

   Section 3 is devoted to a study of rotating black holes in
five-dimensional gauged supergravity.  Starting from a general class
of charged rotating black holes found in \cite{d5gauge1,d5gauge2}, we
obtain expressions for their mass, angular momentum, charges,
temperature, angular velocity and electrostatic potentials, and we
confirm that the first law of thermodynamics holds.  We then turn to a
consideration of the constraints placed by supersymmetry.  We outline
the derivation of a supersymmetric bound, which allows us to deduce
how the parameters of the solutions must be chosen in order to obtain
supersymmetric configurations.  We then investigate the global
structure and regularity of these supersymmetric configurations.  The
classification is quite rich.  Roughly speaking, the supersymmetric
solutions fall into two classes, which we shall call A and B.

    The solutions in class A generically preserve $\ft12 $ of the
${\cal N}=2$ supersymmetry.  Their classification resembles in some
respects that of the BMPV black holes.  For certain ranges of the
parameters, the solutions are free of CTC's, but
there is a naked singularity.  For other ranges of the parameters,
 CTC's are present.  Inside the region of CTC's (\ie inside the
time machine), there is what formally looks like an event horizon,
with imaginary temperature and area.  However, this is just a
coordinate singularity; the spacetime can be closed off to give a
non-singular metric at this point, at the expense of making a {\it
real} periodic identification of the time coordinate.  We refer to 
the ``horizon'' in this case as a pseudo-horizon.  The
identification of the time coordinate is the explanation of the
previously puzzling observation that the temperature becomes imaginary
in this case. The resulting spacetime resembles the repulson case of
the BMPV solution.  However, in the BMPV case, there was no need to
identify the time periodically.  In the present case, the
identification of the time coordinate requires that we impose a
Josephson-type condition, in order that the fermions in the
gauged supergravity be well defined.

   There is an intermediate case of critical rotation, which is 
perhaps the most
interesting, and appears to be an entirely new type of {\it
topological soliton} solution.  It is a completely non-singular
globally stationary spacetime, with no horizon, defined on the product
of time with a spatial manifold having the topology of an $\R^2$
bundle over $S^2$.  The metric is globally hyperbolic, and completely
free of CTC's. It differs from the recently-discovered bubble or
droplet solutions, which are globally static, charged but
non-rotating, and topologically trivial \cite{liluma,chlupo}.  By
contrast, this new solution is topologically non-trivial, and carries
angular momentum.  We have checked that the new solution has zero
entropy, but satisfies the first law of thermodynamics with the $TdS$
term making no contribution.  In some cases, the soliton solutions
are quantum-mechanically consistent.

   The five-dimensional supersymmetric solutions of type B generically
preserve $\ft14$ of the ${\cal N}=2$ supersymmetry, and generically
have CTC's.  There is either a naked singularity inside the region of
CTC's (\ie inside the time machine), or there is a formal horizon
inside the time machine, at which the the spacetime closes off to give
a geodesically-complete repulson type solution with periodic time.  In
addition there are two further special cases.  In one, the CTC's are
confined inside a Killing horizon to give a supersymmetric black hole
that is regular outside and on the horizon.  This is the solution
obtained in \cite{gutrea1,gutrea2}.  The other special case arises when the
horizon and the boundary of the time machine coincide.  This is a regular
soliton solution, with no horizons and no CTC's, analogous to the
soliton solution we have found amongst the type A solutions.

   Section 4 is devoted to a study of rotating black holes in
seven-dimensional gauged supergravity.  We obtain the mass, angular
momentum, charges, temperature, angular velocity and electrostatic
potentials of a general class of 2-charge black holes constructed in
\cite{d7gauge}, and confirm that the first law of thermodynamics is
satisfied.  The supersymmetric cases are again investigated using the
supersymmetry bound, which allows us to determine the parameters for
which supersymmetry is possible.  Again, there are two classes, which
we shall refer to as class A and class B.  Solutions in class A generically
preserve $\ft38$ of the ${\cal N}=2$ supersymmetry. If the angular momentum
is positive, all such solutions have naked singularities inside a time
machine.   This continues to be true if the angular momentum is
negative, up to a critical value below which we have a repulson-like
behaviour inside a time machine.  The critical case arises when the
boundary of the repulson coincides with the boundary of the time machine,
and again we obtain a completely non-singular topological soliton
type solution.

   The seven-dimensional supersymmetric solutions of type B generically
preserve $\ft18$ of the ${\cal N}=2$ supersymmetry.  Generically, these
solutions have naked singularities or naked CTC's.  However, we find for
special choices of the parameters that we can obtain completely
regular black holes, with CTC's occurring only inside the horizon.  These new
solutions are seven-dimensional analogues of the five-dimensional rotating
black holes found in \cite{gutrea1,gutrea2}.  We
also find a non-singular topological soliton for another special choice 
of the parameters.

  Section 5 is devoted to rotating black holes in four-dimensional
gauged supergravity.  We obtain the thermodynamic quantities for the
general class of 4-charge black holes (with charges set pairwise equal)
that were obtained in \cite{d4gauge}, and we confirm that the first
law is satisfied.  We obtain the conditions under which the solutions are
supersymmetric, and find new regular supersymmetric rotating black holes.

   In section 6, we apply some of the ideas and concepts developed
earlier to some related solutions of current interest in
five-dimensional ungauged supergravity.  These solutions describe
black holes in a G\"odel universe.  The background is supersymmetric,
but contains CTC's.  We show that if the black hole is sufficiently
large that the horizon formally enters the region of CTC's, then a
repulson phenomenon takes place in which the space compactifies, and
time must be identified periodically.  We conclude section 6 with a
further discussion of the homogeneous G\"odel background spacetime
itself.  We show that if one wishes to make identifications to
compactify the spatial sections, it is necessary to make the time
periodic also, thereby obtaining a complete non-singular compact
spacetime.

\section{Global Considerations}

\subsection{Causality}

   Consider a Lorentzian metric on some spacetime $M$
 with two commuting Killing vector fields
$\p_t$ and $\p_\psi$
Assume also that $\psi \in (0,{4 \pi \over k} ]$, $ k \in \Bbb N$.
The metric  may be given the form
%%%%%
\ben
ds ^2 =g_{tt} dt ^2 +2g_{t\psi} (d \psi +\omega) dt +
g_{\psi \psi}  (d \psi +\omega)^2 + g_{ab} dy^a dy^b\,.\label{genmet}
\een
%%%%%
We have assumed that all $dt dy^a$ terms vanish and that all $d\psi
dy^a$ terms may be absorbed in $\omega$ which is a one form on the
quotient manifold $Q$ spanned by the coordinates $y^a$. We shall not
assume that $g_{ab}$ is necessarily positive definite everywhere on
$Q$.  The metric will be of Lorentzian signature if
%%%%%
\ben
g_{tt} g_{\psi \psi} - g_{t\psi}^2   <0\,, \label{det}
\een
%%%%%
and  $g_{ab}$ is positive definite. The boundary of the  region
for which (\ref{det}) holds, \ie the null hypersurface
$H \in M$  on which
%%%%%
\ben
 g_{tt}\,  g_{\psi \psi} - g_{t\psi}^2 =0\,, \label{hor}
\een
%%%%%
is a {\it Killing Horizon}, that is, a stationary null surface,
invariant under time translations.  Thus it can be traversed by timelike
curves in only one direction.  The null generator is
%%%%%
\ben
l=\p_t + \Omega \p_\psi\,,
\een
%%%%%
where
%%%%%
\ben
\Omega = -{ g_{t\psi}  \over g_{\psi \psi} } \Big |_H
\een
%%%%%
The region $D \subset  Q$ outside the horizon with boundary
$\p D=H$, is called in black holes physics
the {\it domain of outer communication}.

   We shall assume that the domain of outer communication contains a
region for which $g_{tt} <0$, so that $\p _t$ is timelike and the
metric is stationary in this region.  The {\it ergo-region} consists
of points in $Q$ for which $g_{tt} >0$.  The boundary $\p D$ is called
the {\it ergo-sphere}, although topologically there is no reason why
it should have the topology of a sphere.  It is a general result that
if the metric admits a Killing spinor, then the associated Killing
vector $\p_t$ must always be non-spacelike, $g_{tt} \le 0$, and and
hence ergo-regions cannot exist. This also means that if one has a
Killing horizon in this case, and if $g_{\psi \psi} >0 $, then we must
have $g_{t \psi} =0$ on the horizon, which implies that $\Omega=0$,
\ie the horizon is non-rotating.  This was first seen in the case of
the BMPV black holes \cite{gaumyetow}.

   If $g_{\psi \psi} >0$, and $t$ is well-defined and non-periodic in
some region, there cannot be closed timelike curves in that region,
because we can use $t$ as a {\it time function} such that $t$ is
non-decreasing along any future-directed timelike curve. In fact
%%%%%
\ben
g^{\mu \nu} \p_\mu t\, \p_\nu t = -{g_{\psi \psi} \over g_{t\psi}^2 -
g_{tt} g_{\psi \psi } }\,.
\label{norm}
\een
%%%%%
Thus the level sets $t={\rm constant}$ will
be spacelike  hypersurfaces, and traversable only once,  as long
as we are outside both the the horizon and $g_{\psi \psi} >0$.
Note that the coordinate function $t$ certainly becomes singular
on the Killing horizon, because the right-hand side of (\ref{norm})
becomes infinite there.

   Conversely, if the condition $g_{\psi \psi} > 0$ is violated, then
the metric will certainly admit closed timelike curves, or CTC's, in
the region $C \subset Q $ for which $g_{\psi \psi}<0$.  Its boundary
$\p C$, given by $g_{\psi \psi} =0$, will admit a closed null curve,
or CNC. We shall refer to $C$ as the {\it time machine} and its
boundary as the {\it velocity of light surface} or VLS.  As long as
$g_{t \psi} \ne 0$ when $g_{\psi \psi}=0$, the metric will remain
non-singular there.  Moreover the velocity of light surface will be a
{\sl timelike hypersurface}, and so timelike curves may cross into the
time machine, and emerge from it, possibly earlier than when they
entered.  Because it is a timelike surface, the VLS is a distinct concept
from that of a Cauchy horizon, which is necessarily a null hypersurface.

    There are three cases of interest.
\begin{itemize}

\item The VLS is inside the horizon
\item The VLS is on the horizon
\item The VLS is outside the horizon

\end{itemize}

   Of special interest is the last case. The metric may be re-expressed as
%%%%%
\ben
ds ^2 = \bigr (g_{tt} - {g_{t\psi} ^2 \over g_{\psi \psi}}\bigl)
dt^2 + g_{ab}dy^a dy^b +
g_{\psi \psi} (d \psi + \omega + {g_{t\psi}  \over g_{\psi \psi} } dt )^2\,.
\een
%%%%%
The first two terms should now be a positive definite metric
on the quotient $ M/SO(2)$ of the manifold by the action
of shifts in $\psi$. The orbits of $\p_\psi$  are timelike
inside the time machine ($g_{\psi \psi} <0 $ ).   The spacetime 
then comes to an end inside the time machine,
where coefficient of $dt^2$ in the metric on the quotient
%%%%%
\ben
\bigr (g_{tt} - {g_{t\psi} ^2 \over g_{\psi \psi}}\bigl) dt^2
  + g_{ab}dy^a dy^b
\een
%%%%%
vanishes.  We shall refer to this as the {\it pseudo-horizon}. 
In order that the spacetime be non-singular
on the pseudo-horizon, it will in general be necessary to identify the 
coordinate $t$ with an appropriate (real) period.

   The time period is easily seen to be related
to the formal expression for the surface gravity $\kappa$ of the 
pseudo-horizon, namely
%%%%%
\ben
\kappa ^2= \nabla _\mu L \nabla ^\mu L\,,
\een
%%%%%
with
%%%%%
\ben
 -L^2 = g_{tt} + 2 g_{t \psi} \Omega + g_{\psi \psi}  \Omega ^2.
\een
%%%%%
In fact one finds that   $\kappa^2$ is negative.
Formally, this suggests an imaginary temperature $T= { 2 \pi \over \kappa}$.

   In other words {\it ${2 \pi \over |\kappa|}$
is the period in real time, rather than the imaginary time period of the
usual  case}. For the same reason, the area of the horizon
is purely imaginary in this case.

   Of course it can be that the surface gravity $\kappa$ actually
vanishes on the horizon. Then, there is no need to make an
identification of the time coordinate. This was found to occur in the
case of the BMPV limiting solution.  The resulting object has been
referred to as a {\it repulson } \cite{gibher}.  An
examination of geodesics in that example showed that they could not
penetrate the horizon.

\subsubsection{BMPV Black Hole}

   To see this in detail
consider the metric of the supersymmetric BMPV black hole \cite{BMPV}, 
%%%%%
\ben
ds ^2 =
- \Delta^2 \bigl (dt + {a m \over 2 \Delta r^2 } \sigma _3 \bigr )^2
+ { dr^2 \over \Delta^2 } +
{r^2 \over 4} \bigl ( \sigma_1 ^2 + \sigma_2^2+ \sigma _3^2  \bigr)\,,
\label{bmpvmet}
\een
%%%%%
with $a$ a constant and
%%%%%
\ben
\Delta =  1-{m \over r^2 }\,.
\een
%%%%%
The energy $E$  and angular momentum $J_R$  are
%%%%%
\ben
E={3 \pi m \over 4} \,\qquad J_R=- {\pi a m \over 2}
\een
%%%%%
The horizon is located at
$r= R_H= \sqrt m$ and the boundary of the time machine is located
at $r= r_L= (a m ) ^{ 1\over 3} $.

   In the over-rotating case $a  > m $, the boundary of the time machine
lies outside the horizon.
The surface gravity of the horizon vanishes and the area is
%%%%%
\ben
A= 2 \pi^2 \sqrt{r_H^6 - r_L^6 }\,.
\een
%%%%%
In this over-rotating case
the area becomes imaginary because $\p_\psi$ is timelike.

   If $\Delta_L= 1 - a^2 m^2/r^6$, then the metric (\ref{bmpvmet}) may be
written as
%%%%%
\be
ds_5^2 = -\fft{\Delta^2\, dt^2}{\Delta_L} + \fft{dr^2}{\Delta^2} +
\ft14 r^2\, \Delta_L\, \Big(\sigma_3 - \fft{2am\Delta}{r^4\, \Delta_L}\,
    dt\Big)^2 + \ft14 r^2\, (\sigma_1^2 + \sigma_2^2)\,.\label{bmpvmet2}
\ee
%%%%%
In the over-rotating case, with $\Delta_L$ negative inside the time
machine, the first two terms
in (\ref{bmpvmet2}) define a two-dimensional positive-definite metric.
The horizon, $\Delta=0$, is an infinite proper radial distance away.
Defining a new radial coordinate $y=\Delta^{-1}$, the metric near the repulson
at the horizon becomes approximately
%%%%%
\be
ds_5^2 \sim \ft14 m\, \fft{dy^2}{y^2} + \fft{dt^2}{y^2\, |\Delta_L(r_H)|}
   + \ft14 m\, [\sigma_1^2 + \sigma_2^2 - |\Delta_L(r_H)|\, \sigma_3^2)\,,
\ee
%%%%%
where $\Delta_L(r_H)=1-a^2/m <0$.  The first two terms are the standard
metric of negative curvature on the upper half plane $\Im(z)>0$, where
%%%%%
\be
z= \fft{2t}{\sqrt{m\, |\Delta_L(r_H)|}} + \im\, y\,.
\ee
%%%%%
If one maps the upper half plane into the unit disc, then the repulson
corresponds to a single point on the conformal boundary circle.

   In the marginal case where $\Delta$ and $\Delta_L$ go to zero
simultaneously, \ie $r_H=r_L$, one obtains a singular object of zero
area.

\subsection{Frames rotating and non-rotating at infinity}\label{frames}

     When discussing the thermodynamics of rotating black holes in
anti-de Sitter backgrounds, it is convenient to define energies and
angular momenta with respect to a frame that is non-rotating at
infinity.  If, however, the system is supersymmetric, there is another
natural frame, determined by the Killing vector associated with the
Killing spinor.  Consider, to begin with, the case of no black hole;
\ie pure AdS$_n$ satisfying $R_{\mu\nu} = - (n-1)\, g^2\, g_{\mu\nu}$.
There is a static Killing vector field $K=\del/\del t$, where $t$ is
the usual static time coordinate of AdS$_n$.  However, the Killing
spinors give rise to everywhere non-spacelike Killing vectors
%%%%%
\be
K_{\pm} = \fft{\del}{\del t} + g \sum_i \ep_i\, 
           \fft{\del}{\del\phi_i}\,,\label{kpm}
\ee
%%%%%
where $\phi_i$ are the canonical angular coordinates in $[(n-2)/2]$
orthogonal spatial 2-planes, whose periods are $2\pi$, and the signs
of the $\ep_i=\pm 1$ the summation may be chosen independently for
each $i$.  

   If one were to adapt the AdS$_5$ metric to a Killing vector $K_+$
with a spinorial square root, one would obtain a stationary metric. If
one projects the metric orthogonal to the Killing field $K_+$, \ie if
one takes the quotient with respect to the one-parameter subgroup of
isometries generated by $K_+$, one would obtain the
(Einstein-K\"ahler) Bargmann metric on $SU(2,1)/U(2)$.  In the
pseudo-orthonormal frame adapted to this rotating frame, the Killing
spinors are independent of time.  In a pseudo-orthonormal frame
adapted to a static Killing field, the spinors depend on time in a
periodic fashion, being proportional to $\exp (i{g \over 2} t)$.
Similarly they will have an angular dependence proportional to
$\exp(\pm \im{\psi\over 2})$.  Analogous remarks apply in other spacetime
dimensions.

    In general, there will always be a Killing vector asymptotic to
$K$, but if black holes are present, it may well have an ergo-region
in which $g(K,K)\equiv g_{\mu\nu}\, K^\mu K^\nu$ becomes positive.
There will also always be Killing vectors $K_{\pm}$, and again in
general there is no reason why $g(K_{\pm}, K_{\pm})$ should be
everywhere non-positive.  However, if the solution preserves ${\cal N}$
supersymmetries, \ie it has ${\cal N}$ linearly-independent Killing
spinors, for which
%%%%%
\be
\bar\epsilon\, \gamma^\mu\, \ep\, \del_\mu = K_{\pm}\,,
\ee
%%%%%
for some choices of the $\pm$ signs in (\ref{kpm}).  We shall refer to
Killing vectors that are associated with Killing spinors in this way
as having a spinorial square root.  The spinorial square root need not
be unique.  In general the sum of two such Killing vectors will not
have a spinorial square root.  Note that even in a non-extremal
solution, we can define $K_\pm$ as the Killing vectors that
can be expressed asymptotically in terms of spinorial square roots, since
any asymptotically AdS spacetime admits Killing spinors in the
asymptotic region near infinity.

    Consider one of these special Killing vector fields.  It is
future-directed, rotating ``at the speed of light'' at infinity, and
is nowhere spacelike.  It coincides, on the Killing horizon, with the
null generator, and hence, for any supersymmetric black hole, the
angular velocities on the horizon are $\pm g$. In other words, a
supersymmetric black hole in an AdS background rotates with angular
velocity $\pm g$ with respect to a frame that is non-rotating at
infinity.

\subsection{Geodesics}

    In order to study geodesic completeness, we need to be able to
solve for the geodesics.  All of the metrics considered in this paper
take the form (\ref{genmet}), with
%%%%%
\ben
g_{ab} dy^a dy^b = g_{rr} dr ^2 + A^2(r) g_{ij}dx^idx^j\,,
\een
%%%%%
where
%%%%%
\ben
\omega= \omega_i(x)  dx^i\,,
\een
%%%%%
with $g_{tt} ,g_{t \psi}$ and $g_{\psi \psi}$ being functions
only of the coordinate $r$.

    The Hamilton-Jacobi equation for neutral particles is
%%%%%
\bea
&&{ 1 \over A^2 (r) } g^{ij} (\p_i S -\omega _i \p _\psi S)
(\p_j S - \omega _j \p_\psi   S )
+ g^{rr} \p_r S \p_r S  +  g ^{tt } \p_t S \p_ t S \nn\\
&&+
2 g^{t \psi} \p_t S \p_\psi S + g^{\psi \psi} \p_\psi S \p _\psi S
  = - m^2 .
\eea
%%%%%
We separate the equation by setting
%%%%%
\ben
S=-Et + j \psi +W(r) + F(x^i)\,,
\een
%%%%%
and find that
%%%%%
\ben
 { K^2 \over A^2(r) }  +  g^{rr} ( {d W \over dr})^2    +  E^2 g ^{tt }
-2 E j g^{t \psi}    + j^2  g^{\psi \psi}      = - m^2 ,
\een
%%%%%
with the constant $K$ satisfying
\ben
g^{ik} ( \p_i F - j \omega _i  ) (\p_k  F- j \omega_k)= K^2\,.
\een
%%%%
The radial equation is then given by
%%%%%
\bea
m {dr \over  d \lambda } \!\!&=& \!\!g^{rr} \p_r W \label{radial}\\
\!\! \!\!\!  &=& \!\!
\pm\Bigl (  { g^{rr}  \over g_{t\psi}  ^2 - g_{tt }g_{\psi \psi} }
 \Bigl )^ {1 \over 2}   \Bigl
( E^2 g _{\psi \psi }  +  2  E j  g_{t \psi} +
 j^2 g_{t t }  -  ( g_{t\psi}  ^2 - g_{tt } g_{\psi \psi}    )
 ( m^2 + { K^2 \over A^2(r) })   \Bigr )^{ 1 \over 2}  \,.\nn
\eea
%%%%%

   The equation for $x^i$ is determined in terms of the motion
of a fictitious charged particle in a magnetic field $\omega_i$
moving on the manifold whose coordinates are $x^i$:
%%%%%
\ben
m {dx^i \over d \lambda}= { g^{ik} \over A^2(r) }  ( \p_k F -j \omega_k )\,.
\een
%%%%%
The motion in time and angle is given by
%%%%%
\bea
m {dt \over d \lambda} &=& { 1 \over g_{t\psi}  ^2 - g_{tt } g_{\psi \psi}}
\bigl(  E g_{\psi \psi}   + j  g_{t \psi} \bigr)\,,\nn\\
m {d\psi  \over d \lambda}&=& { 1 \over g_{t\psi}  ^2 - g_{tt } g_{\psi \psi}}
\bigl(  - j g_{tt}  - E  g_{t \psi} \bigr )\,.
\eea
%%%%%
If the time machine is outside the horizon, then from (\ref{radial}) it follows
that particles with $j=0$ cannot enter the time machine.  However, by
taking $E$ and $j$ such that $2g_{t\psi} E j $ sufficiently
large and positive, one can find
geodesics where $dr/d\lambda$ is still negative as $r$ approaches $r_L$,
and so such geodesics can enter the time machine.

    For particles with charge $e$ moving in a potential $A_\mu$, the
the geodesics are governed by the Hamilton-Jacobi equation with the
replacement
%%%%%
\ben
\p_\mu S \rightarrow \p_\mu S + e A_\mu\,,
\een
%%%%%
so that
%%%%%
\ben
m{d x^\mu \over d \lambda} = g^{\mu \nu} \bigl( \p_\nu S + e A_\nu )\bigr )\,.
\een
%%%%%
If we assume that
\ben
A =  \phi (r)  dt + \chi (r)  (d \psi +\omega _i dx^i)\label{a1form}
\een
the discussion above will go through as before with the replacements
%%%%%
\ben
E \rightarrow E - e \phi(r)\,,\qquad j \rightarrow j + e \chi(r)\,.
\een
%%%%%
One may think of $\phi$ an electrostatic potential
and $\chi$ as a magnetostatic potential.

\subsection{Energetic Considerations}

   From (\ref{radial}), it follows that
a  neutral particle may only cross  the horizon if
%%%%%
\ben
E^2 g_{\psi\psi}  + 2 E j g_{t \psi}  + j^2 g_{tt}   \ge  0
\een
%%%%%%
on the horizon. Bearing in mind that the quadratic form
%%%%%
\ben
\Omega^2 g_{\psi \psi} + 2 \Omega g_{t \psi} + g_{tt}
\een
%%%%%
has a pair of coincident roots $\Omega= -g_{t\psi}/g_{\psi\psi}$
on the horizon, we must have
%%%%%
\ben
g_{\psi\psi}\, (E-\Omega j)^2 >0\,.\label{ejsq}
\een
%%%%%
If the time machine is inside the horizon then $g_{\psi\psi}>0$, and hence,
assuming $E>0$, we must have
%%%%%
\be
E - \Omega\, j > 0\,.\label{ej2}
\ee
%%%%%
Since $E$ and $j$ give the small increment in the energy and angular
momentum of the black hole, one may regard this inequality as a
statement of the second law of thermodynamics.  If, on the other hand,
the time machine is outside the horizon, then $g_{\psi\psi}<0$ on
the horizon, and (\ref{ejsq}) can never be satisfied.  This means that
the repulson is a barrier preventing penetration by timelike geodesics.

   In the case of charged particles, one defines the electrostatic
potential of the horizon by $\Phi=l^\mu\, A_\mu$, where $l^\mu\p_\mu
= \p_t + \Omega \p_\psi$ is the  null generator of the horizon.  This
implies
%%%%%
\ben
\Phi = \bigl( \phi +\Omega \chi \bigr) \Big | _H \,.
\een
%%%%%
The second law for the case of a time machine inside the horizon
then generalises from (\ref{ej2}) to become
%%%%%%
\ben
E - \Omega j - \Phi  e > 0\,.
\een
%%%%%

\subsection{Quantisation Conditions}

   The vector potential (\ref{a1form}) in general becomes singular at the
horizon, as can be seen from its norm
%%%%%
\be
A_\mu\, A_\nu\, g^{\mu\nu} = \fft{\chi^2\, g_{tt} + \phi^2\, g_{\psi\psi}
       - 2\phi\chi\, g_{t\psi}}{g_{tt}\, g_{\psi\psi} -g_{t\psi}^2}\,.
\ee
%%%%%
It is then necessary to make a gauge transformation
%%%%%
\be
A\rightarrow \wtd A= A - d(\phi(r_+)\, t)\,.
\ee
%%%%%
A field $\Psi$ with charge $e$ will thus suffer a
gauge transformation
%%%%
\be
\Psi\rightarrow \exp(\im\, e\, \phi(r_+)\, t)\,\Psi\,.\label{psitran}
\ee
%%%%%

   If the coordinates $t$ is periodic, with period $\Delta t$, this 
can lead to the quantisation condition
%%%%%
\be
\fft{e\, \phi(r_H)\, \Delta t}{2\pi}\in \Z \,.\label{jcond}
\ee
%%%%%
If the original gauge potential $A$ and the new potential $\wtd A$ are
both needed in order to define the connection over the entire space, then the
$U(1)$ transition function given by (\ref{psitran}) will be
well-defined only if it is a periodic function of $t$ such that
(\ref{jcond}) is satisfied.  Typically, the need for the two gauge 
patches arises if the potential $A$ falls off sufficiently rapdidly at
infinity but $\wtd A$ does not.

   The quantisation condition (\ref{jcond}) is unfamiliar since one
does not normally consider time to be periodic.  An analogous
condition arises when considering the time-dependent Josephson effect,
and so we propose calling it the Josephson quantisation condition.

\section{Rotating Black Holes and Supersymmetric Limits in 
Five-Dimensional Gauged Supergravity}

\subsection{Black-Hole Thermodynamics in Five Dimensions}

   Our starting point is the non-extremal charged rotating black holes
in five-dimensional gauged supergravity, which were found in
\cite{d5gauge1,d5gauge2}.  Specifically, we shall focus on the
solutions obtained in \cite{d5gauge2}, which describe rotating black
holes with the two rotation parameters set equal, and with three
independent electric charges carried by the three commuting $U(1)$
gauge fields in the maximal $SO(6)$ gauged theory.  The relevant part
of the supergravity Lagrangian is given by
%%%%%
\be
e^{-1}\, {\cal L} = R - \ft12{\del\vec\varphi}^2 -
  \ft14\sum_{i=1}^3 X_i^{-2}\, {(F^i)}^2  + 4 g^2 \,
  \sum_{i=1}^3 X_i^{-1} + \ft1{24} |\ep_{ijk}|\, \ep^{\mu\nu\rho\sigma\lambda}
  F^i_{\mu\nu}\, F^j_{\rho\sigma}\, A^k_{\lambda}\,,\label{d5lag}
\ee
%%%%%
where $\vec\varphi=(\varphi_1,\varphi_2)$, and
%%%%%
\be
X_1= e^{-\fft1{\sqrt6}\varphi_1 -\fft1{\sqrt2} \varphi_2}\,,\qquad
X_2= e^{-\fft1{\sqrt6}\varphi_1 +\fft1{\sqrt2} \varphi_2}\,,\qquad
X_3 = e^{\fft2{\sqrt6}\varphi_1}\,.
\ee
%%%%%

   The solutions in \cite{d5gauge2} are
characterised by five non-trivial parameters, associated with the
mass, the angular momentum, and the three electric charges.  As
presented in \cite{d5gauge2}, the metrics were written with an
additional trivial parameter called $\gamma$, which characterises the
asymptotic rotation rate as measured from infinity.  For convenience, we
shall set $\gamma=0$, which means that the metric is asymptotically
non-rotating.  The solution of \cite{d5gauge2} is then given by
%%%%%
\bea
ds_5^2 &=& (H_1 H_2 H_3)^{1/3}\,
 \Big\{ -\fft{r^2\, Y}{f_1}\, dt^2 + \fft{r^4}{Y}\, dr^2
+ \fft{f_1}{4r^4\, H_1 H_2 H_3}\,
  (\sigma_3  -\fft{2 f_2}{f_1}\, dt)^2\nn\\
&&\qquad\qquad\qquad  + \ft14 r^2 (\sigma_1^2 + \sigma_2^2) \Big\}\,,\nn\\
A^i &=& \fft{2m}{r^2\, H_i}\, \Big(s_i\, c_i\, dt +\ft12 a \,
(c_i\, s_j\, s_k
      - s_i \, c_j\, c_k)\, \sigma_3\Big) \,,\nn\\
X_i &=&  \fft{R}{r^2\, H_i}\,,\ \ \ i=1,2,3\,,\label{d5sol}
\eea
%%%%%
where
%%%%%
\bea
H_i &=&  1 + \fft{2m\, s_i^2}{r^2}\,,\nn\\
\sigma_1 + \im\, \sigma_2 &=& e^{-\im\, \psi}\, (d\theta +\im\, \sin\theta\,
  d\varphi)\,,\qquad \sigma_3= d\psi+ \cos\theta\, d\varphi\,,
\eea
%%%%%
and $s_i$ and $c_i$ are shorthand notations for
%%%%%%
\be
s_i\equiv \sinh\delta_i\,,\qquad c_i \equiv \cosh\delta_i\,,\qquad
i=1,2,3\,.
\ee
%%%%%
Note that in the expressions in (\ref{d5sol}) for the vector potentials
$A^i$, the triplet indices $(i,j,k)$ are all unequal: $(i\ne j\ne k\ne i)$.
The functions $f_1$, $f_2$, $f_3$ and $Y$ are given by
%%%%%
\bea
f_1 &=& r^6\, H_1 H_2 H_3 + 2m a^2 r^2 +
     4m^2 a^2[2(c_1 c_2 c_3 - s_1 s_2 s_3)s_1 s_2 s_3 - s_1^2 s_2^2 -
        s_2^2 s_3^2 - s_3^2 s_1^2]\,,\nn\\
f_2 &=& 2ma (c_1 c_2 c_3 - s_1 s_2 s_3) r^2 + 4m^2 a s_1 s_2 s_3\,,\nn\\
f_3 &=& 2ma^2(1+g^2 r^2) + 4 g^2 m^2 a^2
    [2(c_1 c_2 c_3 - s_1 s_2 s_3)s_1 s_2 s_3 - s_1^2 s_2^2 -
        s_2^2 s_3^2 - s_3^2 s_1^2]\,,\nn\\
Y &=& f_3 + g^2 r^6 H_1 H_2 H_3 + r^4 - 2m r^2\,.
\eea
%%%%%
(Here we have renamed the parameters $\mu$ and $\ell$
  in \cite{d5gauge2} as $2m$ and $a$ respectively.)

    We wish to calculate the conserved charges corresponding to the
energy $E$, the angular momentum $J$ and the three electric charges $Q_i$.
The electric charges are easily calculated from
%%%%%
\be
Q_i = \fft1{8\pi}\, \int_{S_3}\,( X_i^{-2}\,  {*F^i}
         -\ft12 |\ep_{ijk}|\, A^j\wedge F^k)\,,
\ee
%%%%%
where the integration is performed over the 3-sphere at
infinity.\footnote{The prefactor $1/(8\pi)$  rather than $1/(4\pi)$
results from our normalisation ${\cal L} \sim
 R - \ft14 F^2$ rather than $R-F^2$ in the Lagrangian.}
The angular momentum can be derived from the Komar integral
%%%%%
\be
J = \fft1{16\pi}\, \int_{S^3} {*dK}\,,
\ee
%%%%%
where $K= K_\mu\, dx^\mu$, and $K^\mu\, \del_\mu = \del/\del\psi$
is the rotational Killing vector conjugate to the angular momentum.
The calculation of the energy $E$ is trickier because the analogous
Komar integral for the relevant timelike Killing vector diverges
in an asymptotically AdS background, and requires a delicate and
somewhat ambiguous regularisation.  As discussed in \cite{gibperpop},
one of the simplest ways of calculating the energy in such a situation
is to integrate up the first law of thermodynamics, which in this
case reads\footnote{The coefficient 2 in the angular momentum contribution
comes from $\Omega_1\, dJ_1 + \Omega_2\, dJ_2$, with the two angular
momentum contributions being equal.}
%%%%%
\be
dE = T\, dS + 2\Omega\, dJ + \sum_i \Phi_i\, dQ_i\,,\label{firstlaw}
\ee
%%%%%
where $T$ is the Hawking temperature, $S$ is the area of the horizon,
$\Omega$ is the angular velocity of the horizon relative to the
frame that is non-rotating at infinity, and $\Phi_i$ are the
electrostatic potentials on the horizon.  Since all the other quantities
are relatively easily calculated without ambiguity, this provides a
convenient way to compute the energy.\footnote{There are unambiguous methods
available for directly computing the energy of an asymptotically AdS
spacetime, such as the conformal definition of the conserved energy
given by Ashtekar, Magnon and Das (AMD) \cite{ashmag,ashdas}.  
In fact in \cite{gibperpop}
the energies of rotating AdS black holes in all dimensions were calculated
both from the integration of the first law, and by the AMD procedure, and the
two results were shown to be identical.}  

   For the rotating black holes (\ref{d5sol}), we have
%%%%%
\bea
S&=& \ft12 \pi^2\, \sqrt{f_1},\qquad
T= \fft{Y'}{4\pi\, r\, \sqrt{f_1}}\,,\qquad \Omega= \fft{2f_2}{f_1}\,,\nn\\
\Phi_i &=& \fft{2m}{r^2\, H_i}\, \Big(s_i c_i - \ft12 a\, \Omega\,
   (c_i s_j s_k - s_i c_j c_k)\Big)\,,
\eea
%%%%%
with all quantities evaluated on the outer horizon $r=r_+$ where
$Y(r)$ has its largest root.  Note that $\Omega$ here is the angular
velocity measured with respect to the Killing vector $\del/\del\psi$.
In terms of azimuthal coordinates $\phi_1$ and $\phi_2$, with
canonical periods $2\pi$, in a generic solution with independent
angular velocities $\Omega_1$ and $\Omega_2$ in the two orthogonal
2-planes, one has $\psi=\phi_1+\phi_2$, $\varphi=\phi_1-\phi_2$, and
hence $\del/\del\psi = \ft12\del/\del\phi_1 + \ft12 \del/\del\phi_2$,
and hence $\Omega= \ft12(\Omega_1+\Omega_2)$.  
When the two angular velocities are equal, as in our case, one therefore has
$\Omega=\Omega_1=\Omega_2$.

    After calculations of some complexity, we
can integrate the first law (\ref{firstlaw}) to obtain the energy $E$.
Ours results for the various conserved quantities in this case are:
%%%%%
\bea
E &=& \ft14 m\, \pi\, (3 + a^2 g^2 + 2s_1^2 + 2s_2^2 + 2s_3^2)\,,\nn\\
J &=& \ft12 m\, a\, \pi\, (c_1 c_2 c_3 - s_1 s_2 s_3)\,,\label{d5ejq}\\
Q_i &=& \ft12 m\, \pi\, s_i c_i\,.\nn
\eea
%%%%%
It should be emphasised that the fact one obtains an exact form on the
right-hand side of (\ref{firstlaw}), and hence that it can be
integrated, is a somewhat non-trivial result, which, in its own right,
provides a significant test of the validity of the first law for these
black hole solutions.  

   If one turns off the charges, by setting
$s_i=0$, these expressions reduce to ones obtained for
five-dimensional rotating AdS black holes in \cite{gibperpop}.  If the
three charges are set equal, $s_i=s$, the expressions reduce to the
ones found in \cite{madros}.

\subsection{Supersymmetric Limits and the Supersymmetric 
                    Bound}\label{d5bogsec}

   The algebra of the supercharges ${\cal Q}$ in five-dimensional
${\cal N}=2$ gauged supergravity, \ie the $U(1)$ gauged minimal 
supergravity, is given by $\{{\cal Q},{\cal Q}\}=
\{\overline{{\cal Q}},\overline{{\cal Q}}\}=0$
and
%%%%%
\be
M_\equiv
\{ {\cal Q}, \overline{{\cal Q}} \} = \ft12 
\, J_{AB}\, \gamma^{AB} + Z\,,
\label{susy5}
\ee
%%%%%
where $\gamma^{AB}$ are generators of $SO(4,2)$ in the $8\times 8$ 
spin representation, and the supercharge $Q$ is a Weyl spinor of 
$SO(4,2)$.   $J_{05}$ is the energy, and $J_{ij}$ for $1\le i\le 4$
correspond to angular momenta.\footnote{The corresponding
four-dimensional boundary theory is usually referred to as an ${\cal
N}=1$ superconformal theory.}   We are
interested in the case where there are two rotation parameters $J_1$
and $J_2$, describing rotations in the $J_{12}$ and $J_{34}$ planes.
We have
%%%%%
\be
J_{05}= g^2\, E\,,\qquad J_{12} = g^3\, J_1\,,
  \qquad J_{34} = g^3\, J_2\,,\qquad
 Z = g^2\,  \sum_i Q_i\,.
\ee
%%%%%
The \bog matrix $g^{-2}\, M$ has in general four distinct eigenvalues
associated with four complex Weyl eigenspinors of the given chirality.
These are given by
%%%%%
\bea
\lambda &=& E  + g J_1 + g J_2 - \sum_i Q_i\,,\nn\\
\lambda &=& E  - g J_1 - g J_2 - \sum_i Q_i\,,\nn\\
\lambda &=& E  + g J_1 - g J_2 + \sum_i Q_i\,,\nn\\
\lambda &=& E  - g J_1 + g J_2 + \sum_i Q_i\,,\label{4evs}
\eea
%%%%%
each with degeneracy 1.   

   We could instead have used a notation where the supercharges were 
Majorana.  This would lead to each of the four eigenvalues in (\ref{4evs})
being associated with two real spinors.  

   If the two angular momenta are set equal, $J_1=J_2=J$, we obtain
%%%%%
\bea
\lambda &=& E + 2 g\, J - \sum_i Q_i \qquad \hbox{once}\,,\label{emq0}\\
\lambda &=& E - 2 g\, J - \sum_i Q_i \qquad \hbox{once}\,,\label{emq00}\\
\lambda &=& E + \sum_i Q_i \qquad \hbox{twice}\,.\label{epq0}
\eea
%%%%%%
The vanishing of one of these eigenvalues is associated with the 
occurrence of supersymmetries (see \cite{gibhulwar,ferpor}).
We shall refer to the situation where states have vanishing eigenvalues
given by (\ref{epq0}) as supersymmetric configurations of type A.  By
contrast, the situation where the eigenvalue in (\ref{emq0}) or
in (\ref{emq00}) vanishes will be associated with supersymmetric 
configurations of type B.   Note that the  the identification
of the R-charges is confirmed  by both the local supersymmetry
transformation rules  and the Witten-Nester identity \cite{gibhulwar}.  

   The discussion above is given in terms of the ${\cal N}=2$ theory,
which can be embedded in the maximal ${\cal N}=8$ theory, for which
there are four complex Weyl supercharges and the R-symmetry is
$SU(4)$.  The \bog matrix now carries R-symmetry indices, and in general
has 16 distinct eigenvalues for Weyl eigenspinors of the given chirality.
These eigenavlues are
%%%%%
\be
\lambda= E \pm g j_1 \pm g J_2 \pm Q_1\pm Q_2 \pm Q_3\,,
\ee
%%%%%
where, according to the convention chosen for the chirality, the
total number of minus signs must be either always odd, or always even.
Again, one could instead use a notation where the supercharges were 
Majorana.  In this case each of the 16 eigenvalues would be associated
with two Majorana eigenspinors.

   Taking our results (\ref{d5ejq}) for the conserved charges of the
non-extremal rotating AdS black holes in five dimensions, we find that
%%%%%
\bea
 E + 2 g\, J - \sum_i Q_i \!\!&=&\!\! \fft{m\pi}{4}
 (1+ a g \, e^{\delta_1+\delta_2+\delta_3})(e^{-2\delta_1} +
              e^{-2\delta_2} + e^{-2\delta_3} + ag\, e^{-\delta_1-\delta_2
               -\delta_3})\,,\nn\\
 E - 2 g\, J - \sum_i Q_i\!\! &=&\!\! \fft{m \pi}{4}
 (1- a g \, e^{\delta_1+\delta_2+\delta_3})(e^{-2\delta_1} +
              e^{-2\delta_2} + e^{-2\delta_3} - ag\, e^{-\delta_1-\delta_2
               -\delta_3})\,,\nn\\
E + \sum_i Q_i \!\!&=&\!\! \fft{m \pi}{4} (e^{2\delta_1} + e^{2\delta_2} +
         e^{2\delta_3} + a^2 g^2)\,.\label{d5evs}
\eea
%%%%%
We see that vanishing of these quantities can be achieved in the
following cases:
%%%%%
\bea
 E + 2 g\, J - \sum_i Q_i &=& 0\qquad \hbox{if}\quad
     e^{\delta_1+\delta_2+\delta_3}= -\fft{1}{ag}\,,\\
 E - 2 g\, J - \sum_i Q_i &=& 0\qquad \hbox{if}\quad
     e^{\delta_1+\delta_2+\delta_3}= \fft{1}{ag}\,,\\
E + \sum_i Q_i  &=& 0\qquad \hbox{if}\quad  m\rightarrow 0\,,\quad m\,
   e^{-2\delta_i} = 2q_i\,,\label{epq}
\eea
%%%%%%
where in the last case the limit is achieved by sending the $\delta_i$
parameters to $-\infty$ whilst keeping the $q_i$ fixed.\footnote{The
other ostensibly supersymmetric cases that one might think could
arise, such as if $(e^{-2\delta_1} +e^{-2\delta_2} + e^{-2\delta_3} +
ag\, e^{-\delta_1-\delta_2 -\delta_3})=0$ in the first line of
(\ref{d5evs}), are in fact spurious, and are not associated with
supersymmetric limits of the black holes we are discussing.  This can
be verified by checking the explicit supergravity transformation
rules, or, more easily, by checking to see whether either of the
``spinorial square root'' Killing vectors $K_+$ or $K_-$ introduced in
section \ref{frames} has a manifestly non-positive norm, as would have
to be the case if there existed a Killing spinor in the background.   
Evidently the solutions corresponding to one of these spurious roots
have singularities that violate the assumptions under which one can
deduce supersymmetric from a vanishing eigenvalue of the \bog matrix.} 

   The positivity of the left-hand side
of the supersymmetry  algebra (\ref{susy5}) implies quite generally
 that the AdS and R-charges
live in a convex cone invariant
under the adjoint action of the product
of the R-symmetry group and the anti-de-Sitter group.
This invariant cone may be rotated
to lie in the maximal torus, which is spanned
by the energy, angular momenta and U(1) charges.
The boundary of the cone consists of states with
some degree of supersymmetry. The  boundary is a  stratified
set consisting of faces, edges, etc., with the strata of smaller
dimension consisting of increasingly larger
numbers of zero eigenvalues, \ie of increasing amounts of supersymmetry
\cite{gagihuto}.

    In this $D=5$ example, truncated to a single R-charge $ \sum_i
 Qi$, supersymmetry allows only states labelled by points in the four
 dimensional vector space with coordinates $E$, $g J_1$, $g J_2$ and
$\sum_i Qi$, which satisfy the four inequalities
%%%%%
\bea
E+gJ_1 + gJ_2  -\sum_i Q_i &\ge& 0\,,\nn\\
E -gJ_1 - gJ_2  -\sum_i Q_i &\ge& 0\,,\nn\\
E +gJ_1 - gJ_2  + \sum_i Q_i &\ge& 0\,,\nn\\
E -gJ_1 + gJ_2  + \sum_i Q_i &\ge& 0\,. 
\eea
%%%%%
This is a convex cone  in $\R^4$, 
with four faces given by the four hyperplanes through
the origin, whose equations are given by the saturation of the inequalities 
above, \ie when one of the four eigenvalues vanishes.
The base of the cone is obtained by setting
$E={\rm constant} >0$ in these inequalities, and
is easily seen to consist of  points in $\R^3$, 
inside or on a regular tetrahedron. This has four faces, 
corresponding to the vanishing of one of the four eigenvalues. 
If $J_1=J_2=J$, then the third and fourth faces intersect on an edge 
of enhanced supersymmetry for which
\be
%%%%%
E + \sum_i Qi=0\,.
\ee
%%%%%
Thus states on the faces are what was earlier called type B.
The states on the edges are what was earlier called type A.

%\subsection{Global structures of BPS solutions}

\subsection{$E + \sum Q_i=0$: Case A}
\label{sec:d5ks}

      This solution preserves $\ft12$ of the ${\cal N}=2$ supersymmetry,
since there are two zero eigenvalues of the \bog matrix in this case
(see equation (\ref{epq0})).
     The supersymmetric condition $E+\sum Q_i=0$ can be satisfied
by taking $m\rightarrow 0$ and $\delta_i\rightarrow -\infty$,
whilst leaving $q_i=\ft12 m\, e^{-2\delta_i}$ fixed.  If
It is also necessary to scale the parameter $a$ according to
%%%
\be
a=\ft12 \alpha \sqrt{\fft{2m}{q_1 q_1 q_3}}\,,
\ee
%%%%
so that the gauge fields remains finite.
After taking $m\rightarrow 0$, the solution then becomes
%%%%%
\bea
ds^2&=& -\ft14 r^2\, \fft{V}{B}\, dt^2 + \fft{dr^2}{V} +
B\, (\sigma_3 + f dt)^2 + \ft14 R^2 (\sigma_1^2 + \sigma_2^2)\,,
\nn\\
A^i &=& -\fft{1}{r^2\, H_i}(q_i\, dt - \ft12\alpha \sigma_3)\,,
\label{d5ks}
\eea
%%%%
where
%%%
\bea
V=\fft{r^4 + g^2 r^6\, H_1 H_2 H_3 - g^2\alpha^2}{r^4\, (H_1 H_2 H_3)
^{1/3}}\,,\qquad
B=\fft{r^6H_1H_2H_3 -\alpha^2}{4 r^4(H_1H_2H_3)^{2/3} }\,,\nn\\
f=-\fft{2\alpha\,r^2}{r^6 H_1H_2H_3 -\alpha^2}\,,\qquad
R^2=r^2 (H_1H_2H_3)^{1/3}\,,\qquad
H_i=1 + \fft{q_i}{r^2}\,.
\eea
%%%%
 From (\ref{d5ejq}), the energy, angular momentum and the charges
are given by
%%%
\be
E=\ft14\pi (q_1 + q_2 + q_3)\,,\qquad J=\ft14\pi\,\alpha\,,\qquad
Q_i=-\ft14\pi\,q_i\,,
\ee
%%%%
Note that we have
%%%
\be
g(K,K) =-\fft{V r^2}{4B} + B f^2=
 -\fft{g^2 r^2 H_1H_2 H_3 + 1}{(H_1 H_2 H_3)^{2/3}}\,,
\label{d5ksgkk1}
\ee
%%%
where $K$ is the asymptotically non-rotating timelike Killing
vector introduced in section \ref{frames}, and which is given by
$K=\del/\del t$ in this case.  The function $g(K,K)$ is
negative definite provided $r^2 H_i>0$.  Thus the Killing vector $K$
is nowhere spacelike, but since it is non-rotating at infinity it does
not have a spinorial square root.  It is straightforward to calculate
$g(K_+,K_+)$ for the Killing vector $K_+=\del/\del t + 2g\, \del/\del\psi$
that does have a spinorial square root; we find
%%%%%
\be
g(K_+,K_+) = -\fft{V r^2}{4B} + B (f + 2g)^2=
-\fft{(r^2 + g\alpha)^2}{r^4 (H_1 H_2 H_3)^{2/3} }\,.
\label{d5ksgkk2}
\ee
%%%
This is indeed manifestly non-positive.

   The solutions (\ref{d5ks}) were obtained previously in \cite{klemm2}.
Here we examine further their global properties.  Note that the range
of the coordinate $r$ is $r^2 + q\ge 0$, with $r_0^2=-q$ being
the singularity, where $q={\rm min}\{q_1,q_2,q_3\}$, with $q_i>0$.
Thus there must exist a VLS where $B=0$ for non-vanishing rotation,
inside which there are CTC's. If $V$ is never zero for all $r^2 + q\ge 0$,
the solution has a naked singularity.  This happens when there is
no rotation (without CTC's), or under-rotation (with CTC's).

     Since the right-hand side of the (\ref{d5ksgkk1}) is negative
definite, it follows that if there is a Killing horizon where $V=0$,
then in general $B$ must be negative, implying CTC's.  This
in fact rules out any regular black hole solution.  The only
possibility for the solution not to have a naked CTC is when both
$B$ and $V$ approach zero simultaneously.\footnote{In the identity
(\ref{d5ksgkk2}), it appears that the right-hand side could be
zero if we choose to have $\alpha=-r_0^2/g$. This implies
that $H_1 H_2 H_3=0$ at $r=r_0$ for the existence of a horizon.
This however forces $B$ to be divergent at $r=r_0$.}
This can occur when $\alpha^2 = q_1 q_2 q_3$.  We shall now
discuss this case in detail.

\subsubsection{Critical rotation: $\alpha^2=q_1q_2q_3$}

    In this case, there is neither an horizon nor a time machine,
and the radial coordinate $r$ runs from 0 to $\infty$.  Near to $r=0$,
the metric becomes
%%%
\bea
ds^2 &\sim &-g^2 (q_1 q_2 q_3)^{1/3}\, dt^2 +
\fft{q_1 q_2+ q_1 q_3 + q_2 q_3}{4 (q_1 q_2 q_3)^{2/3}}\, r^2 (\sigma_3
-\fft{2(q_1 q_2 q_3)^{1/2}}{q_1 q_2 +q_1 q_3 + q_2 q_3} dt)^2\nn\\
&&+ \fft{(q_1 q_2 q_3)^{1/3}\, dr^2}{g^2 (q_1 q_2 + q_1 q_3 + q_2 q_3)}
+ \ft14 (q_1 q_2 q_3)^{1/3} (\sigma_1^2 + \sigma_2^2)\,,
\label{smallr}
\eea
%%%
A conical singularity at $r=0$ is avoided if the charges are chosen to
satisfy the condition
%%%
\be
\fft{g (q_1 q_2 + q_1 q_3 + q_2 q_3)}{k\, \sqrt{q_1 q_2 q_3}} =1\,,
\qquad k   = \fft{g\, \alpha}{k}\, \sum_i \fft1{q_i} 
   \in \Z_+\label{lumpchargecon}
\ee
%%%%
with $\psi$ having period $\Delta\psi=4\pi/k$ for any strictly 
positive integer $k$.

   Setting $dt=0$ in the metric (\ref{d5ks}) with $\alpha^2 = q_1 q_2
q_3$ gives a four-dimensional metric that is everywhere
positive-definite, except at the coordinate singularity at $r=0$.
From the behaviour near $r=0$ given in (\ref{smallr}), we have seen
that the metric may be extended to give a complete non-singular metric
on an $\R^2$ bundle over $S^2$, with the coordinates $r$ and $\psi$
parameterising the $\R^2$, and $\theta$ and $\phi$ are coordinates on
the $S^2$ base.  Such $\R^2$ bundles are characterised topologically
by the single integer $k$, which gives the Chern number of the bundle.
If $k=1$ we obtain the spin bundle of $S^2$, \ie the Taub-BOLT manifold
\cite{pagebolt}, and if $k=2$ we obtain
the tangent bundle of $S^2$, \ie the Eguchi-Hanson manifold 
\cite{begipapo}. \footnote{Of course the {\it metrics} on these manifolds,
given in \cite{pagebolt} and \cite{egha}, differ from those we are finding
in this paper, but the {\it topologies} are the same.  The correct topology
for the Eguchi-Hanson metric was first discussed in \cite{begipapo}.}
The total five-dimensional spacetime is topologically a product of the
four-dimensional manifold with the real line.  It is a spin manifold if
and only if the $\R^2$ bundle over $S^2$ is a spin manifold, and this
is the case if and only if $k$ is even.  If $k$ is odd, the manifolds 
admit a spin$^c$ structure, \ie one may consistently couple spinors
provided they carry an appropriate $U(1)$ charge with respect to a
suitable Maxwell gauge field \cite{hawpop}.  A brief discussion of the 
spin$^c$ structure of the Taub-BOLT metric is given in appendix A. 

   It is interesting to note that these topological solitons, which as we
have seen have zero temperature, still satisfy the
first law of thermodynamics, namely, we have
%%%%%
\be
dE = 2 \Omega\, dJ + \sum_i \Phi_i dQ_i\,.
\ee
%%%%%

    We now turn to the regularity of the gauge fields.  
    The gauge potentials in (\ref{d5ks}) are well-defined for all
$r>0$, with, in particular
%%%%%
\be
A^i_\mu\, A^i_\nu\, g^{\mu\nu} \sim \fft{g^2\, q_1 q_2 q_3 - q_i^2}{g^2 r^6}
\ee
%%%%%
as $r\longrightarrow\infty$.  They are, however, singular at $r=0$,
diverging as
%%%%%
\be
A^i_\mu\, A^i_\nu\, g^{\mu\nu} \sim
\fft{(q_1 q_2 q_3)^{5/3} }{ q_i^2 (q_1 q_2 + q_2 q_3 + q_3 q_1)\, r^2}
\ee
%%%%%
at small $r$.  We can define new gauge potentials, ${A^i}'$ and ${A^i}''$
that are well-defined at $r=0$, $\theta=0$, and at $r=0$, $\theta=\pi$ 
respectively, by means of the gauge transformations
%%%%%
\bea
A^i&\longrightarrow& {A^i}'= A^i + \fft{\alpha}{2 q_i}\, (d\psi+d\phi)\,,\nn\\
A^i&\longrightarrow& {A^i}''= A^i + \fft{\alpha}{2 q_i}\, (d\psi-d\phi)\,.
\eea
%%%%%
These potentials are, however, not well-defined at infinity, since
one has
%%%%%
\be
{A_\mu^i}'\, {A_\nu^i}'\, g^{\mu\nu} \sim 
 \fft{2\alpha^2\, g^2(1-\cos\theta)}{r^2\, \cos^2\theta}\,,\qquad
{A_\mu^i}''\, {A_\nu^i}''\, g^{\mu\nu} \sim
  \fft{2\alpha^2\, g^2(1+\cos\theta)}{r^2\, \cos^2\theta}
\ee
%%%%%
as $r$ approaches infinity.  Thus
we must necessarily cover the manifold in three gauge patches, and 
associated with these are quantisation conditions arising when we
require the quantum well-definedness of the gauged supergravity's
fermionic wave functions, which are all gauged with respect to the
graviphoton $U(1)$ connection $\ft12 g\, (A^1_\mu+A^2_\mu+A^3_\mu)$.
The wave functions in the 
three gauge patches are related by the transition functions
%%%%%
\bea
U_1 &=& \exp(\ft{\im}{2}\, g\, \alpha\, \sum_i \fft{1}{q_i}\, \phi)=
      e^{\ft12 \im\, k\, \phi}\,,\nn\\
U_2 &=& \exp(\ft{\im}{4}\, g\, \alpha\, \sum_i \fft{1}{q_i}\, (\psi+\phi))=
      e^{\ft14 \im\, k\, (\psi+\phi)}\,,\nn\\
U_3 &=& \exp(\ft{\im}{4}\, g\, \alpha\, \sum_i \fft{1}{q_i}\, (\psi-\phi))=
      e^{\ft14 \im\, k\, (\psi-\phi)}\label{tranu}\,.
\eea
%%%%%
(The second equalities on each line follow from
(\ref{lumpchargecon}).)  

    If the spacetime manifold admits a spin
structure, these transition functions must be appropriately
single-valued.  Thus $U_1$, which governs the behaviour of the 
graviphoton $U(1)$ bundle
restricted to the $S^2$ base, must satisfy $U_1(\phi+2\pi) =U_1(\phi)$,
whilst $U_2$ and $U_3$, which govern the behaviour of the graviphoton
$U(1)$ bundle
restricted to the $\R^2$ fibres, must satisfy $U_2(\psi + 4\pi/k) =
U_2(\psi)$ and $U_3(\psi + 4\pi/k) = U_3(\psi)$.  As we discussed
earlier, the condition for having a spin structure is that $k$ be
even.  Thus although $U_1$ is indeed then single-valued, we see that
$U_2$ and $U_3$ are not, since $\psi$ has period $4\pi/k$.  

    If, on the other hand, the spacetime is not a spin manifold, which
happens if $k$ is odd, then the fermions must be sections of a
spin$^c$ bundle, as we discussed above.  This implies that
the transition functions should satisfy $U_1(\phi+2\pi)=-U_1(\phi)$, 
and $U_2(\psi+4\pi/k) =-U_2(\psi+4\pi/k)$, 
$U_3(\psi+4\pi/k) =-U_3(\psi+4\pi/k)$.  The minus signs in these relations
precisely cancel the minus signs coming from the transition functions in
the spin connection that are responsible for the absence of an ordinary
spin structure \cite{hawpop}.

    The argument given above, based on the transition functions for
the graviphoton $U(1)$ bundle over the spatial manifold, itself an
$\R^2$ bundle over $S^2$, with connection $\ft12 g \, \sum_i A^i$, may
be expressed more concisely checking the Dirac quantisation
condition, which for a spin structure is
%%%%%
\be
{g \over 4 \pi} \int_C \sum_i F^i \in {\Bbb Z}\,,\label{diracint}
\ee
%%%%%
and for a spin$^c$ structure is
%%%%%
\be
{g \over 4 \pi} \int_C \sum_i F^i \in {\Bbb Z} +\ft12\,,\label{diracint2}
\ee
%%%%%
where the integrals are taken over all relevant 2-cycles $C$.   Because the
spatial manifold is non-compact, the cycles in our case consist of the
$S^2$ base at $r=0$, and the non-compact cycle corresponding to the 
$\R^2$ fibre over any point of the $S^2$ base.  These give
%%%%%
\be
\fft{g}{4\pi}\int_{S^2} \sum_i F^i = -\ft12 k\,,\qquad
\fft{g}{4\pi}\int_{\R^2} \sum_i F^i = -\ft12\,.\label{fints}
\ee
%%%%%
When $k$ is even, in which case the spacetime is a spin manifold, 
we see that first integral indeed satisfies the spin condition 
(\ref{diracint}), but the second integral can never satisfy this
condition.  When $k$ is odd, in which case the spacetime is a 
spin$^c$ manifold, we see that both the integrals in (\ref{fints}) 
satisfy the spin$^c$ conditions (\ref{diracint2}).  Thus we conclude, in
agreement with our discussion of the transition functions above, that
the fermions of the supergravity multiplet are quantum-mechanically
well-defined in the case that $k$ is odd, but not when $k$ is even.  
It should be emphasised, however, that in all cases the bosonic 
soliton solutions are completely regular.

\subsubsection{General rotation: $\alpha^2 \ne q_1 q_2 q_3$}

   In this case, if the parameters $\alpha$, $q_1$, $q_2$ and $q_3$
lie in appropriate ranges, there is a Killing horizon at $r^2=r_+^2$,
the largest value of $r^2$ where $V$
vanishes, and at which $B<0$.  The functions metric $V$ and $B$ depend
on $r$ only via $r^2$, and so the values of $r^2$ at which they vanish
might in fact be negative.  The metric (\ref{d5ks}) itself is real and
of Lorentzian signature for values of $r^2$, including negative ones,
such that $r^2 + q_i>0$ for all $i$.  With this understanding, the VLS
occurs at $r^2=r_L^2> r_+^2$, where $B(r_L)=0$. In the region between,
$r_+^2 \le r^2 < r_L^2$, the metric has CTC's.  Since the Killing
horizon at $r=r_+$ lies inside the VLS, it is what we have defined
earlier as a pseudo-horizon.

   Note that in order to avoid a conical singularity at $r^2=r_+^2$, the time
coordinate $t$ must be identified with the (real) period
%%%
\bea
\Delta t &=& \fft{8\pi \sqrt{-B(r_+)}}{r_+\, V'(r_+)}\nn\\
  &=&\fft{2\pi}{g(2 + g^2 r^2 (H_1H_2 + H_1H_3 + H_2H_3))}\Big|_{r=r_+}\,.
\label{tperiod}
\eea
%%%%
The metric is then geodesically complete, with $r^2$ ranging from $r^2=r_+^2$
to $\infty$.  The periodic coordinates $t$ and $\psi$ play the role
of time in different regions.  Outside the time machine, $r^2>r_L^2$,
$t$ is the time coordinate, whilst inside the time machine, $r^2<r_L^2$,
it is $\psi$ that is timelike.  

    The solution we have obtained has the topology $\R^2\times S^3$,
where the angular coordinate $t$ and the radial variable $r2$
parameterise the $\R^2$ factor, while the Euler angles $\theta$,
$\phi$ and $\psi$ parameterise the $S^3$.  Strictly speaking, in order
that the Euler coordinates not break down near $r^2=r_+^2$, one must
introduce a shifted Euler coordinate $\psi'= \psi + f(r_+)\, t$.  With
this choice, the Euler angles $\theta$, $\phi$ and $\psi'$ have only
the usual singularities on $S^3$, but they are globally defined for
all $r^2$ and $t$.  Because $r^2=r_+^2$ is merely the centre of polar
coordinates on the $\R^2$ factor, radial geodesics cannot reach values
of $r^2$ less than $r_+^2$, and so the solution is of repulson type.

   We now turn to an examination of the quantum mechanical consistency
of these solutions.  We begin by noting that the gauge potentials
are not simultaneously well-defined at both the pseudo-horizon and at
infinity.  The gauge potentials given in (\ref{d5ks}) satisfy
%%%%%
\be
A^i_\mu\, A^i_\nu\, g^{\mu\nu} = 
  \fft{\alpha^2\, H_i(1 + g^2 r^2 H_1 H_2 H_3 H_i^{-2}) -
       (H_i-1)^2 r^6\, H_1 H_2 H_3 H_i^{-1} }{r^6 H_i (H_1 H_2 H_3)^{2/3}\, 
            V}\,,\label{Asq}
\ee
%%%%%
which can be seen to diverge on the pseudo-horizon at $r^2=r_+^2$, 
where $V$ vanishes.
On the other hand, at large $r^2$ we see from (\ref{Asq}) that the $A^i$
are non-singular, with the asymptotic behaviour
%%%%%
\be
 A^i_\mu\, A^i_\nu\, g^{\mu\nu} \sim \fft{\alpha^2\, g^2 - q_i^2}{g^2\, r^6}
\,.
\ee
%%%%%
We can define gauge-transformed potentials 
%%%%%
\be
{A^i}' = A^i + c_i\, dt\,,\label{tgauge}
\ee
%%%%%
which are regular on the pseudo-horizon at $r^2=r_+^2$, with the constants
$c_i$ given by
%%%%%
\be
c_i\equiv  \fft{2 q_i + \alpha\, f(r_+)}{2 r_+^2\, H_i(r_+)}\,.
\label{cidef}
\ee
%%%%%
These potentials are however not pure gauge at infinity, \ie
%%%%%
\be
\oint {A^i}' \ne0\,,
\ee
%%%%
where the integral is taken over the closed timelike loop parameterised 
by $t$ at infinity.  This can also be seen from
%%%%%
\be
{A^i}'_\mu\, {A^i}'_\nu\, g^{\mu\nu} \sim -\fft{c_i^2}{g^2\, r^2}\,,
\ee
%%%%%
as $r^2\longrightarrow\infty$. 

   Requiring the quantum consistency of the fermion wave functions under
the above gauge transformation implies that the phase factor
%%%%%
\be
U= e^{\ft12\, \im\, g\,\sum_i c_i\,  t}
\ee
%%%%
should be single-valued.  With the period of $t$ given by (\ref{tperiod}),
this implies the Josephson quantisation condition
%%%
\be
\fft{g\,\Delta t}{8\pi\, r_+^2}\Big(\fft{2q_1 + \alpha\,f}{H_1(r_+)} +
\fft{2q_2 + \alpha\,f}{H_2(r_+)}+ \fft{2q_3 +\alpha\,f}{H_3(r_+)}
\Big) =  \td n \in \Z\,,
\ee
%%%%
and hence from (\ref{tperiod})
%%%
\be
\td n = 
\ft12 + 
    \fft{1}{4 + 2g^2 r^2 (H_1H_2 + H_1H_3 + H_2H_3)}\Big|_{r=r_+}\,.
\label{josephcond}
\ee
%%%
This Josephson condition may be satisfied for appropriate choices of the
parameters (bearing in mind that $r_+^2$ can be negative, provided that
$r_+^2 + q_i>0$ for all $i$ in order to ensure that $R^2$ remains 
positive).

    The Josephson quantisation condition can instead be derived by
integrating the field strengths $F^i$ over the ${\R}^2$ factor, as in
(\ref{diracint}), provided that one is careful in the definition of
the $\R^2$ fibre.  As we mentioned previously, in order to eliminate a
coordinate singularity near the repulson, one must introduce the new
Euler angle $\psi'= \psi + f(r_+)\, t$.  The 1-forms $A^i$ given in
(\ref{d5ks}) now become
%%%%%%
\be
A^i = -\fft1{r^2 H_i}\, \Big( q_i dt - \ft12 \alpha (d\psi'+ \cos\theta
   d\phi) -\ft12 \alpha\, f(r_+) dt\Big)\,.
\ee
%%%%%
We integrate $F^i=dA^i$ over the 2-surface defined by taking $\theta$, 
$\phi$ and $\psi'$ to be constant, yielding
%%%%%
\be
\fft{g}{4\pi}\, \int_{\R^2} \sum_i F^i =\ft12 + 
    \fft{1}{4 + 2g^2 r^2 (H_1H_2 + H_1H_3 + H_2H_3)}\Big|_{r=r_+}\,.
\ee
%%%%%

    We have seen that in general, a time machine implies the periodic
identification of the time coordinate $t$, which implies an
appropriate restriction of parameters to achieve quantum consistency.
A special case arises, in which $t$ does not need to be periodically
identified, if $V$, considered as a function of $r^2$, has a double
root.  This can be achieved with the following choice of parameters:
%%%
\bea
&&\alpha^2 = q_\3 + \ft12 q_\2 r_0^2 - \ft12 r_0^6\,,\qquad
g^2 = -\fft{2r_0^2}{q_\2 + 2q_\1 r_0^2 + 3 r_0^4}\,,\nn\\
&&q_\1 \equiv q_1 + q_2 + q_3\,,\qquad
q_\2\equiv q_1q_2 + q_1q_3 + q_2q_3\,,\qquad
q_\3\equiv q_1q_2q_3\,.\label{paraspec}
\eea
%%%%
Note that here $r_0^2<0$, with the constraint $r_0^2 + q_i>0$ for
all $q_i$.  The function $V$ is then given by
%%%
\be
V=\fft{g^2(r^2-r_0^2)^2 (-q_\2 + 2 r_0^2 r^2 + r_0^4)}{2r_0^2 r^4
(H_1 H_2 H_3)^{1/3}}\,.
\ee
%%%
Here, we present the full solution for the equal charge case:
%%%
\bea
ds^2&=& -\fft{R^2 V}{4B} dt^2 + \fft{dR^2}{V} +
B (\sigma_3 + f dt)^2 + \ft14 R^2 (\sigma_1^2 + \sigma_2^2)
\,,\nn\\
V&=&\fft{(R^2-R_0^2)^2 (1+2g^2 R_0^2 + g^2 R^2)}{R^4}\,,\qquad
B=\ft14 R^2 -\fft{R_0^6 (9g^2 R_0^2 +4)}{16 R^4}\,,\nn\\
f &=& \fft{2\alpha}{BR^2}\Big(1 - \fft{R_0^2(3g^2 R_0^2 +2)}{2R^2}
\Big)\,,\qquad
A_\1^i =-\fft{1}{R^2} (q\, dt - \ft12 \alpha \sigma_3)\,,\nn\\
\alpha^2 &=& R_0^6 + \ft94 g^2 R_0^8\,,\qquad
q=R_0^2 + \ft32 g^2 R_0^4\,.
\eea
%%%%
The solution describes a BMPV type of repulson, for which the $t$ coordinate
is not periodically identified.  One again needs to make gauge 
transformations to define new potentials that are regular on the 
pseudo-horizon; these are the same as in the case with general rotation
that we discussed previously, with the specialisation of parameters 
given in (\ref{paraspec}).  Since there is no periodic identification of
$t$ in this case, no quantum-consistency condition arises.  In fact, this
case corresponds to the situation where the denominator in (\ref{josephcond})
vanishes, and so it is associated with the $\tilde n\rightarrow\infty$ limit
of the solutions with periodically-identified time.

          Finally, for values of the parameters for which $r_+^2+ q_i<0$,
there will be naked singularities.  

\subsection{$E-2g J - \sum Q_i=0$: Case B}
\label{e2gjqsec}

   The cases $E-2g J - \sum Q_i=0$ and $E+2g J - \sum Q_i=0$ are
equivalent, modulo a reversal of the sign of $a$, and the discussion
for the two is equivalent.  We shall consider the former.  As can be
seen from (\ref{emq00}), the \bog matrix will then have one zero
eigenvalue, and so the solution preserves $\ft14$ of the ${\cal N}=2$
supersymmetry.  To satisfy the supersymmetric condition 
%%%%%
\be
e^{\delta_1 + \delta_2 + \delta_3}= \fft1{ag}\,,\label{d5gengrcon}
\ee
%%%%% 
it is most convenient to use it to
express the parameter $a$ in terms of $\delta_i$.  The general
solution is somewhat cumbersome to present, and we shall first discuss
the situation where the three charges are set equal.

   Making the variable changes
%%%
\bea
m=\fft{2\sqrt{g\,\alpha} q^{3/2}}{g\,\alpha -q}\,,\qquad
e^{2\delta} = \sqrt{q/(g\,\alpha)}\,,\qquad
r^2 = R^2 - \fft{(\sqrt{g\,\alpha}-\sqrt{q})^2}{g\,\alpha - q}\,,
\eea
%%%%
the solution becomes
%%%%
\bea
ds_5^2 &=& -\fft{R^2 V}{4B}\, dt^2 + B (\sigma_3 + f\, dt)^2 +
\fft{dR^2}{V} + \ft14 R^2 (\sigma_1^2 + \sigma_2^2)\,,\nn\\
A^i &=& -\fft{1}{R^2}\, (q dt -\ft12 \alpha\,\sigma_3)\,,\nn\\
V&=&1 + g^2 R^2 + \fft{2(q + 2\alpha g)}{R^2} -
\fft{(g\alpha + q)(g\alpha -q)^2 -4\alpha^2}{(g\alpha -q) R^4}
\,,\nn\\
B&=&\ft14 R^2 +\fft{\alpha^2}{(g\alpha - q)R^2} -\fft{\alpha^2}{4R^4}
\,,\nn\\
f&=& \fft{\alpha (g\alpha\,q -q^2 -g\alpha R^2 - 3 q R^2)}{2(g\alpha -q)
R^4 B}\,.
\eea
%%%%
The thermodynamic quantities for this solution are given by
%%%
\be
E=\fft{\pi(3q^2 + 3 g\alpha\,q + 2 g^2\alpha^2)}{4(g\alpha-q)}
\,,\qquad
J=\fft{\pi\alpha (g\alpha + 3q)}{4(g\alpha -q)}\,,\qquad
Q=-\ft14\pi q\,.
\ee
%%%% 
These local solutions were obtained in \cite{d5gauge1,d5gauge2}
using different variables.  Here, we shall discuss in some detail
their global structure.

    First, it is easy
to see that $\del_t$ cannot have a spinorial square root, since
$g_{tt}=g(K,K)$ can become positive.  As discussed previously,
the coordinate transformation $\psi\rightarrow \psi + 2g \, t$,
$t\rightarrow t$, yields $K_+=\del_t +2g \,\del_\psi$ as the Killing
vector with a spinorial square root.  In fact, we have
%%%
\be
g(K_+,K_+) = -\fft{V R^2}{4 B} + B (f + 2g)^2 =
-\fft{(R^2+q -g\alpha)^2}{R^4}\,.
\ee
%%%%%
This equation shows that if there is a Killing horizon at $r=r_+>0$
where $V(r_+)=0$, then $B(r_+)$ must be negative unless either

\begin{itemize}

\item[({\bf i})] The right-hand side of the above equation goes to
zero at the horizon, or

\item[({\bf ii})] $B(r_+)=0$, \ie the VLS coincides
with the horizon.

\end{itemize}
Note that the first case is never possible for the $\ft12$-supersymmetric
solution described in section \ref{sec:d5ks}.  Here, however, it
can be achieved, by choosing the parameters $q$ and $\alpha$ to be
%%%
\be
q=-R_0^2 - \ft12 g^2 R_0^4\,,\qquad
\alpha = -\ft12 g R_0^4\,,
\ee
%%%
whereupon the solution becomes
%%%%
\bea
V&=&\fft{(R^2-R_0^2)^2 (g^2 R^2 + 2g^2 R_0^2 + 1)}{R^4}\,,\qquad
B=\ft14 R^2 + \fft{g^2 R_0^6}{4R^2} - \fft{g^2 R_0^8}{16 R^4}\,,\nn\\
f&=&\fft{1}{4B}\Big( \fft{g R_0^6(g^2 R_0^2 + 2)}{2 R^4} -
\fft{g R_0^4(2g^2 R_0^2 +3)}{R^2}\Big)\,.
\eea
%%%%
In this case the Killing horizon coincides with an event horizon,
and  the VLS occurs inside this horizon.  The solution, which was
obtained in \cite{gutrea1}, describes a supersymmetric black hole that is
regular outside and on the event horizon.

      Alternatively, we can also avoid naked CTC's by considering the
possibility (ii) listed above, where the VLS occurs on the horizon.
This can be achieved by choosing the parameters so that
%%%
\be
q=\fft{\alpha^2}{R_0^4}\,,\qquad
g=\fft{\alpha(\alpha^2 + 3R_0^6)}{R_0^4(\alpha^2 - R_0^6)}\,.
\ee
%%%
Note that in this case, it is more convenient to express
$g$ in terms of $\alpha$.  The metric functions are given by
%%%
\bea
V&=&g^2 (R^2-R_0^2) \Big(1 + \fft{R_0^2 (R_0^6 + \alpha^2) (
R_0^{12} + 6\alpha^2 R_0^6 + \alpha^4)}{\alpha^2 R^2 (3 R_0^6+
\alpha^2)^2}\nn\\
&& \qquad\qquad
+\fft{R_0^4 (R_0^{18}-3\alpha^2 R_0^{12} + 11 \alpha^4
R_0^6 + 7 \alpha^6)}{\alpha^2 R^4 (3R_0^6 + \alpha^2)^2}\Big)\,,\nn\\
%%%%
B&=& \ft14(R^2-R_0^2)(1 + \fft{R_0^2}{R^2} + \fft{\alpha^2}{
R_0^2 R^4})\,,\nn\\
f &=& -\fft{2\alpha^3}{R_0^4(R_0^2 R^4 + R_0^4 R^2 + \alpha^2)}
\,.
\eea
%%%%%
This solution describes a regular soliton, \ie with no horizon and no
CTC's.  To avoid a conical singularity at $R=R_0$, the quantisation condition
%%%
\be
\fft{(2 R_0^6 + \alpha^2)(R_0^6 + 3\alpha^2)}{2R_0^6(R_0^6-\alpha^2)}
 =k
\ee
%%%
must be satisfied, where $k$ is the integer characterising the
topology of the $S^3/\Z_k$ spatial sections.

    Aside from the two special cases enumerated above, all the remaining
solutions will have naked CTC's.  Among these, there are two
possibilities.  One is that the metric has no Killing horizon, and
hence we have a naked singularity cloaked by a VLS.  The other
possibility is that the metric has a pseudo-horizon inside the VLS. This
situation is very much like the $\ft12$-supersymmetric solutions, which we
obtained in section \ref{sec:d5ks}.  In general, it is necessary to
identify the $t$ coordinate periodically,  to avoid a conical
singularity at the horizon.  The geodesics are complete from the
horizon to infinity. Since the pseudo-horizon lies inside the VLS, it is
a repulson.

     A special case of the latter category is where there is a
pseudo-horizon with a double root that lies inside the VLS.  In this case,
regularity on the pseudo-horizon does not require making a periodic
identification of the $t$ coordinate.  The solution is achieved by
choosing the parameters so that
%%%
\be
q=\fft{(9g^2 R_0^2 + 8)(3g^2 R_0^2 +2)}{18 g^2}\,,\qquad
\alpha = -\fft{(9g^2 R_0^2 +4)(3g^2 R_0^2 +2)}{18g^3}\,.
\ee
%%%%
We find that the metric functions become
%%%%
\bea
V&=&\fft{(R^2-R_0^2)^2 (g^2 R^2 + 2g^2 R_0^2 + 1)}{R^4}\,,\nn\\
B&=&\ft14 R^2 - \fft{(9 g^2 R_0^2 +4)^2}{108 g^4R^2} \Big(
1 + \fft{(3g^2 R_0^2 +2)^2}{12g^2R^2}\Big)\,,\nn\\
f &=& -\fft{(9g^2 R_0 + 4)}{648 g^5B}\Big(
\fft{6g^2 (9g^2 R_0^2 + 10)}{R^2} + \fft{(9g^2 R_0^2 + 8)(
3g^2 R_0^2+2)^2}{R^4}\Big)\,,
\eea
%%%%
from which we see that $V$ has a double root at $R=R_0$.  However, the
function $B$ at $R=R_0$ becomes
%%%
\be
B=-\fft{(9g^2 R_0^2 + 2)^2(3g^2 R_0^2 + 4)^2}{1296 g^6 R_0^4}\,.
\ee
%%%
Since this is negative, it implies the occurrence of CTC's outside the
pseudo-horizon.  The structure of this solution is analogous to the BMPV
repulson.

    We now turn to the general situation (\ref{d5sol}) where the
three charges are unequal, but satisfy (\ref{d5gengrcon}),
Using a similar analysis to that for the three equal
charges, we first examine $g(K_+, K_+)$, given by
%%%
\bea
g(K_+,K_+)\!\!\! &=&\!\!\! -\fft{(H_1 H_2 H_3)^{1/3}}{f_1}
\Big(r^2 Y -\fft{(f_2-g f_1)^2}{r^4 H_1 H_2 H_3}\Big)
\label{d5gr3ch}\\
\!\!\!&=&\!\!\! -(H_1 H_2 H_3)^{-2/3} \Big(1 -
\fft{ m (e^{-2\delta_1} + e^{-2\delta_2} + e^{-2\delta_3} -
  e^{-2\delta_1 -2\delta_2-2\delta_3} -2)}{2r^2}\Big)^2\,.\nn
\eea
%%%%
The non-positivity of this quantity implies that in general there is
a pseudo-horizon inside the VLS.  There are two cases where naked CTC's
can be avoided.  One case arises if the horizon coincides with the
the VLS.  This can be achieved by taking
%%%%
\be
m\, g^2=-\fft{2(1 + e^{2\delta_1+2\delta_2} + e^{2\delta_1+2\delta_3} +
e^{2\delta_2+2\delta_3})^2}{(e^{4\delta_1}-1)(e^{4\delta_2}-1)
(e^{4\delta_3}-1) e^{2(\delta_1 + \delta_2 + \delta_3)}}\,.
\ee
%%%%%
Then, the functions $Y$ and $f_1$ have the same root $r=r_0$, given by
%%%
\be
g^2\,r_0^2 =
\fft{2(1 + e^{2\delta_1+2\delta_2} + e^{2\delta_1+2\delta_3} +
e^{2\delta_2+2\delta_3})}{(e^{2\delta_1}+1)(e^{2\delta_2}+1)
(e^{2\delta_3}+1) e^{2(\delta_1 + \delta_2 + \delta_3)}}
\,.
\ee
%%%%
The solution describes a topological soliton, running from $r^2=r_0^2$,
which has spatial sections that are topologically an $R^2$ 
bundle over $S^2$, to AdS$_5$ at infinity.  
The charge parameters must satisfy a quantisation condition in order
to avoid a conical singularity associated with the collapsing
of $\sigma_3$; this condition is given by
%%%
\bea
\!\!\!\!\!\!&&\fft{(e^{2\delta_1} + e^{2\delta_2} + e^{2\delta_3} +
e^{2(\delta_1 + \delta_2 + \delta_3)})(
1 - e^{4\delta_1 + 4\delta_2}- e^{4\delta_1 + 4\delta_3}
- e^{4\delta_2 + 4\delta_3} + 2 e^{4(\delta_1 + \delta_2 +
\delta_3)})r_0}{(e^{4\delta_1}-1)(e^{4\delta_2}-1)
(e^{4\delta_3}-1) e^{2(\delta_1 + \delta_2 + \delta_3)}}\nn\\
\!\!\!\!\!\!&&=k\,,
\eea
%%%
for $S^3/\Z_k$ spatial sections.

   The alternative way to avoid naked CTC's is by ensuring
that the right-hand side of (\ref{d5gr3ch}) is zero at the Killing
horizon.  This requires that the function $Y$ have a root $r^2=r_0^2$ 
given by
%%%%
\be
r_0^2 = \ft12 m (e^{-2\delta_1} + e^{-2\delta_2} + e^{-2\delta_3} -
  e^{-2\delta_1 -2\delta_2-2\delta_3} -2)\,,
\ee
%%%%
which implies that
%%%%
\be
m=\fft{1}{2g^2 \sinh(\delta_1 + \delta_2) \sinh(\delta_1 + \delta_3)
\sinh(\delta_2 + \delta_3)}\,.
\ee
%%%%
Remarkably, $r^2=r_0^2$ is a double root for $Y$.  For the solution to be
free of naked singularities, it is necessary that $r_+^2 H_i >0$ for
all $i=1,2,3$, which places restrictions on the domain of allowed
charge parameters $\delta_i$.  The resulting regular black hole
solutions were previously obtained in \cite{gutrea2}.  The
near-horizon geometry is a direct product of AdS$_2$ and a squashed
3-sphere. The timelike Killing field coincides with an
everywhere-causal Killing field on AdS$_2$, and the system has zero
temperature.  The time machine is located strictly inside the horizon.

    For solutions other than the two examples discussed above, naked
CTC's are inevitable, implying the existence of time machines.
Depending on the parameters, there can be naked singularities, or
regular solutions with $t$ being periodically identified, or else BMPV type
repulsons, where $t$ is not periodic and $Y$ has a double root.
 
\subsection{Lifting to type IIB supergravity}

      All the solutions we considered above can be lifted back
to become solutions of $D=10$ type IIB supergravity.  The reduction
ansatz for the five-dimensional $U(1)^3$ gauged supergravity theory can be
found in \cite{tenauth}.  The metric ansatz is given by \cite{tenauth}
%%%
\be
ds_{10}^2 = \sqrt{\Delta} ds_5^2 + \fft{1}{g^2 \sqrt{\Delta}}
\sum_{i=1}^3 X_i^{-1} \Big(d\mu_i^2 + \mu_i^2
(d\phi_i + g A^i_\1)^2\Big)\,.\label{2bmetric}
\ee
%%%
where $\Delta=\sum X_i\,\mu_i^2$.  With this explicit reduction
ansatz, global properties such as the periods, and CTC's, can be now
addressed from the ten-dimensional point of view.

     In general, the ten-dimensional spacetime is an $S^5$ bundle over
the five-dimensional spacetime, with structural group $U(1)\times U(1)\times
U(1)$, where the $i$'th $U(1)$ acts on $S^5$ by advancing the angle
$\phi_i$.  The requirement that the ten-dimensional metric extend smoothly 
onto a non-singular ten-dimensional manifold leads to the conditions
%%%%%
\be
\fft{g}{2\pi}\, \int_C F^i \in \Z\,,\label{10reg}
\ee
%%%%%
where $C$ is any non-trivial 2-cycle in the five-dimensional spacetime,
since the azimuthal coordinates $\phi_i$ are constrained by the regularity
of the $S^5$ to have periods $2\pi$.   

    For the topological soliton solutions in section 3.3.1, we find
%%%%%
\be
\fft{g}{2\pi} \in \in^{S^2} F^i = -\fft{g\, \alpha}{q_i}\,,
\qquad
\fft{g}{2\pi} \in \in^{\R^2} F^i = -\fft{g\, \alpha}{k\, q_i}\,.
\ee
%%%%%
In view of the fact that all the $q_i$ are necessarily strictly positive,
it follows from (\ref{lumpchargecon}) that the integrals over $\R^2$ can
never satisfy the conditions (\ref{10reg}).  

    For the time machines discussed in section 3.3.2, we find that
for appropriate choices of the parameters we can satisfy the consistency
conditions (\ref{10reg}).  For example, if we set the charges $q_i$ equal
for simplicity, then (\ref{10reg}) will be satisfied if the integer
$\tilde n$ in (\ref{josephcond}) is a multiple of 3.

        It is straightforward to verify that for the $\ft12$
supersymmetric time machine described in section \ref{sec:d5ks}, the
five-dimensional CTC's associated with $\psi$ are no longer CTC's in
ten dimensions, since
%%%
\be
g_{\psi\psi}^{(10)} =\ft12 R^2 \sqrt{\Delta}\,,
\ee
%%%%
which is positive definite. However, CTC's do still exist in ten dimensions.
This can be seen by examining the determinant of the sub-metric involving
the angular coordinates $(\phi_1, \phi_2, \phi_3, \psi\equiv\phi_4)$,
which is given by
%%%%
\be
\det(g_{ij})= \fft{B\mu_1^2\mu_2^2\mu_3^2}{g^6 \Delta}\,.
\ee
%%%
This is negative in the region where $B$ is negative, showing that CTC's are
inevitable in ten dimensions, if they exist in five dimensions (this
was seen in the case of the solutions with CTC's discussed in section
\ref{sec:d5ks} in \cite{calklesab}).

\section{$\!$Rotating Black Holes and Supersymmetric 
Limits in Seven-Dimensional Gauged Supergravity}

\subsection{Black-Hole Thermodynamics in Seven Dimensions}

   Rotating black holes in seven-dimensional gauged supergravity are
somewhat more complicated than those in five dimensions.  One reason
for this is that there is a ``first-order self-duality'' equation for
the 4-form field in the seven-dimensional theory, and this plays a
non-trivial role in the solutions that were obtained in
\cite{d7gauge}.  The relevant part of the Lagrangian for $SO(5)$
gauged supergravity in seven dimensions, in which only the fields that
are non-zero in the solutions are retained, is given by
%%%%%
\bea
{\cal L}_7 &=& R\, {*\oneone} - \ft12 {*d\varphi_i}\wedge d\varphi_i -
  \ft12 \sum_{i=1}^2 X_i^{-2}\, {*F_\2^i}\wedge F_\2^i
    -\ft12 (X_1\, X_2)^2\, {*F_\4}\wedge F_\4\nn\\
&& - 2g^2\, [(X_1\, X_2)^{-4} -8 X_1\, X_2 -4 X_1^{-1} X_2^{-2} -
              4 X_1^{-2}\, X_2^{-1}]\, {*\oneone} \nn\\
&&-g\, F_\4\wedge A_\3
   + F_\2^1\wedge F_\2^2\wedge A_\3\,,\label{d7lag}
\eea
%%%
where
%%%%%
\bea
F_\2^i &=& dA_\1^i\,,\qquad F_\4=dA_\3\,,\nn\\
X_1 &=& e^{-\ft1{\sqrt2}\, \varphi_1 -\ft1{\sqrt{10}}\, \varphi_2}
\,,\qquad
X_2 = e^{\ft1{\sqrt2}\, \varphi_1 -\ft1{\sqrt{10}}\, \varphi_2}\,,
\eea
%%%%%
together with the first-order odd-dimensional self-duality equation
to be imposed after the variation of the Lagrangian.  This condition
is conveniently stated by introducing an additional 2-form
potential $A_\2$, which can be gauged away in the gauged theory, and
defining
%%%%%
\be
F_\3 = dA_\2 - \ft12 A_\1^1\wedge dA_\1^2 - \ft12 A_\1^2\wedge dA_\1^1\,.
\ee
%%%%%
The odd-dimensional self-duality equation then reads
%%%%%
\be
(X_1\, X_2)^2\, {* F_\4} = -2g\, A_\3 - F_\3\,.\label{odddim}
\ee
%%%%%

   The non-extremal rotating black hole solutions found in
\cite{d7gauge} are given by
%%%%%
\bea
ds_7^2 &=& (H_1 H_2)^{1/5}\, \Big[ -\fft{Y\, dt^2}{f_1\, \Xi_-^2}
  + \fft{r^2\, \rho^4\, dr^2}{Y}+ \fft{f_1}{\rho^4\, H_1 H_2 \, \Xi^2}\,
   \Big(\sigma - \fft{2 f_2}{f_1}\, dt\Big)^2 +
   \fft{r^2+a^2}{\Xi}\, d\Sigma_2^2\Big]\,,\nn\\
A_\1^i &=& \fft{2m\, s_i}{\rho^4\,\Xi\, H_i}\, 
(\alpha_i\, dt + \beta_i\, \sigma)
\,, \nn\\
A_\2 &=& \fft{m\, a\, s_1\, s_2}{\rho^4\, \Xi_-^2}\,
    \Big(\fft1{H_1} + \fft1{H_2}\Big)\, dt\wedge \sigma\,,\qquad
 A_\3 = \fft{2m\, a\, s_1\, s_2}{\rho^2\, \Xi\, \Xi_-}\,
           \sigma\wedge J\,,\nn\\
X_i &=& (H_1 H_2)^{2/5}\, H_i^{-1}\,,\qquad
H_i = 1 + \fft{2m\, s_i^2}{\rho^4}\,,\qquad \rho^2= (r^2+a^2)\,,\nn\\
\alpha_1&\equiv& c_1 - \ft12 (1-\Xi_+^2)(c_1-c_2)\,,\qquad
\alpha_2\equiv c_2 + \ft12 (1-\Xi_+^2)(c_1-c_2)\,,\nn\\
\beta_1 &=& -a\, \alpha_2\,,\qquad
\beta_2 = -a\, \alpha_1\,,\nn\\
\Xi_\pm &\equiv& 1 \pm a\, g\,,\qquad \Xi \equiv  1 -a^2\, g^2
                         = \Xi_-\, \Xi_+\,,\label{d7sol}
\eea
%%%%%
where the functions $f_1$, $f_2$ and $Y$ are given by
%%%%%
\crampest{
\bea
f_1 &=& \Xi\, \rho^6\, H_1 H_2 -
     \fft{4\Xi_+^2\, m^2\, a^2\, s_1^2\, s_2^2}{\rho^4} +
 \ft12 m\, a^2\, \Big[4 \Xi_+^2 + 2 c_1\, c_2\,(1-\Xi_+^4)
         + (1-\Xi_+^2)^2\, (c_1^2+c_2^2)\Big]\,,\nn\\
f_2 &=& -\ft12 g\, \Xi_+\, \rho^6\, H_1 H_2 + \ft14 m\, a\, \Big[
  2(1+\Xi_+^4)\, c_1\, c_2 + (1-\Xi_+^4)\, (c_1^2 + c_2^2)\Big]\,,\nn\\
Y &=& g^2\, \rho^8\, H_1 H_2 + \Xi\, \rho^6
+ \ft12 m\, a^2\, \Big[ 4 \Xi_+^2 + 2(1-\Xi_+^4)\, c_1\, c_2
+ (1-\Xi_+^2 )^2 (c_1^2 + c_2^2)\Big]\nn\\
&&-\ft12 m\, \rho^2 \,\Big[
  4\Xi + 2 a^2 g^2 (6+ 8 a g + 3 a^2 g^2)\, c_1\, c_2  -
  a^2 g^2 (2+a g)(2+3a g)(c_1^2+c_2^2)\Big]\,.
\eea
}
%%%%%
The metric $d\Sigma_2^2$ is the standard Fubini-Study metric on $\CP^2$,
and $\sigma=d\psi+ {\cal B}$, where $\ft12 d{\cal B}$ is the K\"ahler form
on $\CP^2$.  The coordinate $\psi$, which has period $2\pi$, lives
on the $U(1)$ fibre of $S^5$ viewed as a $U(1)$ bundle over $\CP^2$.

   In these solutions the three {\it a priori} independent angular
momenta have been set equal.  There are two independent electric
charges, characterised by $s_i=\sinh\delta_i$ (as usual we are also
using the notation $c_i=\cosh\delta_i$).  These charges are carried by
the two 2-form fields of the $U(1)\times U(1)$ subgroup of $SO(5)$.

   We again follow the strategy of \cite{gibperpop} in order to
evaluate the energy $E$ of the black hole solution, by integrating up
the first law of thermodynamics.  First, we note that the time
coordinate $t$ in (\ref{d7sol}) has a non-canonical normalisation, as
measured at infinity, and so we introduce the canonically-normalised
$\td t$, defined by $t=\td t\, \Xi_-$.  The Killing vector $\del/\del
\td t$ is rotating at infinity. In fact we have precisely
$K_+=\del/\del \td t$, where $K_+$ has a spinorial square root near
infinity, as discussed in section \ref{frames}.  We can pass to
non-rotating coordinates by defining a new $U(1)$ fibre coordinate
$\psi'=\psi + g \, \td t$.  After performing these transformations, we
find that the entropy, temperature, angular velocity and electrostatic
potentials on the horizon are given by
%%%%%
\bea
S &=& \fft{\pi^3\, (r^2+a^2)\, \sqrt{f_1}}{\Xi^3}\,,\qquad
T = \fft{Y'}{8 \pi\,r\,  (r^2+a^2)^{3/2}\, \sqrt{f_1}}\,,\qquad
\Omega = g+ \fft{2f_2\, \Xi_-}{f_1} \,,\nn\\
\Phi_i &=& \fft{2m s_i}{\rho^4\, \Xi\, H_i}\, [\alpha_i + \beta_i(\Omega-g)]\,,
\eea
%%%%%
where all functions are evaluated at the outer horizon $r=r_+$ where
$Y(r)$ has its largest positive root.

   Again the evaluation of the electric charges and the angular momentum
is straightforward using integrals over the $S^5$ at infinity.  After
calculations of some complexity, we are then able to obtain the energy
by integration of the first law, which for this case reads
%%%%%
\be
dE= T \, dS + 3\Omega\, dJ + \sum_i \Phi_i\, dQ_i\,.\label{firstlaw7}
\ee
%%%%%
Our results are
%%%%%
\bea
E &=& \fft{m\, \pi^2}{32\Xi^4}\, \Big[ 12 \Xi_+^2(\Xi_+^2 -2)
- 2c_1 c_2\, a^2 g^2\, (21\Xi_+^4 - 20 \Xi_+^3 - 15 \Xi_+^2 - 10\Xi_+
     - 6) \nn\\
&&\qquad\qquad\qquad + (c_1^2+c_2^2)(21 \Xi_+^6 - 62 \Xi_+^5 +
        40 \Xi_+^4 + 13 \Xi_+^2 - 2 \Xi_+  + 6)\Big]\,,\nn\\
J &=& \fft{m\, a\, \pi^2}{16\Xi^4}\, \Big[
 4 a g \Xi_+^2 - 2 c_1 c_2 (2\Xi_+^5 -3 \Xi_+^4 -1)
  + ag (c_1^2+c_2^2)(\Xi_+ +1)(2 \Xi_+^3 - 3 \Xi_+^2 -1)\Big]\,,\nn\\
Q_1 &=& \fft{m\, \pi^2\, s_1}{4\Xi^3}\, \Big[
   a^2 g^2 c_2( 2\Xi_+ +1) - c_1 (2\Xi_+^3 - 3\Xi_+^2 -1)\Big]\,,\nn\\
Q_2 &=& \fft{m\, \pi^2\, s_2}{4\Xi^3}\, \Big[
   a^2 g^2 c_1( 2\Xi_+ +1) - c_2 (2\Xi_+^3 - 3\Xi_+^2 -1)\Big]\,.
\label{d7res}
\eea
%%%%%
In this case, the fact that the right-hand side of (\ref{firstlaw7})
turns out to be an exact form, allowing integration to give the energy 
function $E$, is highly non-trivial, and it provides a striking 
demonstration of the validity of the first law of thermodynamics 
for these seven-dimensional rotating black hole solutions.

\subsection{Supersymmetric Limits and the Supersymmetric Bound}

   The algebra of the supercharges ${\cal Q}$ in seven-dimensional
${\cal N}=2$ gauged AdS supergravity is given by
%%%%%
\be
M\equiv
\{ {\cal Q}, \overline{{\cal Q}} \} = \ft12 J_{AB}\,
          \gamma^{AB}  + Z\,,
\ee
%%%%%
where 
$J_{07}$ is the energy, and $J_{ij}$ for $1\le i\le 6$ correspond
to angular momenta.  If there are three parameters $J_1$, $J_2$ and $J_3$
describing rotations in the $J_{12}$, $J_{34}$ and $J_{56}$ planes, we may
write
%%%%%
\be
J_{07}= g^4\, E\,,\qquad J_{12} = g^5\, J_1\,,
  \qquad J_{34} = g^5\, J_2\,,\qquad J_{56} = g^5\, J_3\,,\qquad
 Z = g^4\, \sum_i Q_i\,.
\ee
%%%%%
(See \cite{ferpor} 
for a discussion of the supersymmetry algebra, and \cite{liumin} for
a discussion of supersymmetric non-rotating AdS black holes in 
seven dimensions.)  The eigenvalues of the \bog matrix 
$g^{-4}\, M$ acting on chiral eigenspinors are given by
%%%%%
\bea
\lambda &=& E + g J_1 - g J_2 -g J_3 - \sum_i Q_i\,,
\qquad \hbox{and 2 cyclic}\,,
 \nn\\
\lambda &=& E + g J_1 -g J_2 -g J_3 +
           \sum_i Q_i\,, \qquad \hbox{and 2 cyclic}\,,
 \nn\\
\lambda &=& E + g J_1 + g J_2 +  g J_3 - \sum_i Q_i\,,\nn\\
\lambda &=& E + g J_1 + g J_2 +  g J_3 + \sum_i Q_i\,,
\eea
%%%%%
where on the first two lines there are two further eigenvalues
corresponding to cycling the $+$ sign onto $J_2$ or $J_3$ instead of
$J_1$.

   If we set the three angular momenta equal, $J_1=J_2=J_3=J$, we get
%%%%%
\bea
\lambda &=& E - g J  - \sum_i Q_i \qquad \hbox{thrice}\,,\nn\\
\lambda &=& E - g J  +  \sum_i Q_i \qquad \hbox{thrice}\,,\nn\\
\lambda &=& E  +3 g J  - \sum_i Q_i \qquad \hbox{once}\,,\nn\\
\lambda &=& E  +3 g J  + \sum_i Q_i \qquad \hbox{once}\,.\label{d7evs}
\eea
%%%%%
Substituting our expressions (\ref{d7res}) for the energy, angular
momentum and charges of the rotating black holes (\ref{d7sol}), we
find that the four cases in (\ref{d7evs}), we find that the vanishing
of $\lambda$ is achieved, respectively, if\footnote{Again, as in the
five-dimensional case, there are spurious roots (which in this case correspond
to quite complicated relations between $a$, $g$ and the $\delta_i$), which do
not correspond to supersymmetric solutions within the class of metrics we 
are considering.} 
%%%%%
\be
e^{\delta_1+\delta_2} =\quad 1 +\fft{2}{ag}\,,\quad
1- \fft{2}{ag}\,,\quad
1- \fft{2}{3ag}\,,\quad
1+ \fft{2}{3ag}\,.
\ee
%%%%%
Thus we are led to supersymmetric limits of the non-extremal rotating
black holes (\ref{d7sol}) when
%%%%%
\be
e^{\delta_1+\delta_2}= 1 \pm  \fft{2}{ag}\,,
\ee
%%%%%
which preserve $\ft38$ of the supersymmetry, and which we designate as
type A.  Also, we obtain supersymmetric limits when 
%%%%%
\be
e^{\delta_1+\delta_2}= 1 \pm \fft{2}{3ag}\,, 
\ee
%%%%%
which preserve $\ft18$ of the supersymmetry, and which we designate as
type B.  As we shall show below, the former in general all have closed
timelike curves, while the latter include a particular case, for a
special choice of the parameters, which gives a perfectly regular
supersymmetric black hole with an horizon.

   The general analysis of the convex cone implied by the positivity
of the \bog matrix eigenvalues described in section \ref{d5bogsec}
applies to this $D=7$ case as well. There are three angular momenta
and so the cone lies in $\R^5$ and is bounded by eight hyperplanes.
If $J_1=J_2=J_3$, then two sets of three of these four-dimensional
hyperplanes or faces intersect on two-dimensional faces of enhanced
supersymmetry.  Again, states on the faces are what we call type B,
while states on the edges are what we call type type A.

\subsection{$E-g\, J - \sum_i Q_i=0$: Case A}\label{egjqsec}

    As we saw previously, this supersymmetry condition is satisfied
provided that 
%%%%%
\be
e^{\delta_1 + \delta_2}=1 + \fft{2}{ag}\,.
\ee
%%%%%  
In general,
the solution describes a naked time machine.  This can be seen by
examining the component of the metric
%%%%%
\bea
g(K_+,K_+) &=&
 \fft{1}{f_1} \Big( \fft{4 f_2^2}{R^4 H_1 H_2 \Xi^2} -
\fft{Y}{\Xi_-^2}\Big)\nn\\
&=& -(1 + ag)^2 R^4 - \ft18m e^{-2\delta_1-2\delta_2} (ag\, e^{\delta_1
+\delta_2} -2 - ag) ( (2 + ag) e^{\delta_1 +\delta_2} - a g)\nn\\
&&\qquad\qquad\times
((e^{\delta_1} - e^{\delta_2})ag + 2 e^{\delta_1})
((e^{\delta_1} - e^{\delta_2})ag + 2 e^{\delta_2})\,,
\eea
%%%%
When the supersymmetry condition is satisfied, the second term vanishes,
and hence
%%%%%
\be
g(K_+,K_+) = -(1+ag)^2 R^4\,,
\ee
%%%%%
which is negative definite for all $R$ outside the singularity at
$R=0$.  Near the horizon, where $Y$ approaches zero, it follows that
$f_1$ must be negative, which implies the occurrence of naked CTC's.

    To discuss the global structure in detail, we first consider the
solution with two equal charges, corresponding to $\delta_1=\delta_2=\delta$.
Making the changes of the variables and coordinates
%%%%
\be
R=\sqrt{\Xi} r\,,\qquad t= \Xi_-\, \td t\,,\qquad
q=\fft{2m\sinh^2\delta}{\Xi^2}\,,\qquad
\alpha = -\fft{a q}{1-a g}\,,\label{d7var1eq}
\ee
%%%%
and imposing the supersymmetry condition $e^{2\delta}= 1+2/(ag)$,
we find that the solution (\ref{d7sol}) becomes
%%%%
\bea
ds^2 &=& H^{2/5} \Big(-\fft{V}{H^2 B} r^2 d\td t^2 +
B (\sigma + f d\td t)^2 + \fft{dr^2}{V} + r^2 d\Sigma_2^2\Big)
\nn\\
A_\1^1 &=& A_\1^2 = \fft{1}{r^4 H}(q d\td t + \alpha \sigma)\,,\quad
A_\2= -\fft{\alpha}{r^4 H}\, d\td t\wedge\sigma\,,\quad
A_\3=-\fft{\alpha}{r^2} \sigma\wedge J\,,\nn\\
X_1&=&X_2=H^{-1/5}\,,\qquad H = 1 + \fft{q}{r^4}\,,\nn\\
V&=&1 + g^2 r^2 H^2 + \fft{2g\, \alpha}{r^4}\,,\quad
B=r^2 - \fft{\alpha^2}{r^8 H^2}\,,\quad
f=\fft1{B}\Big(gr^2 + \fft{\alpha}{r^4H^2}\Big)\,.\label{d7equalc}
\eea
%%%%
Note that
%%%
\be g(K_+,K_+)= H^{2/5} \Big( -\fft{V r^2}{H^2 B} + B f^2\Big) = -
H^{-8/5}\label{d7gkk1} \ee is negative definite.  This rules out the
possibility of a black hole without naked CTC's.  From (\ref{d7res}),
the thermodynamic quantities are now given by
%%%
\be
E=\ft18\pi^2 (4q-5g\alpha)\,,\qquad
J=-\ft18 \pi^2 \alpha\,,\qquad
Q_1=Q_2=\ft14 \pi^2 (q-g\alpha)\,.
\ee
%%%%

        The metric behaviour depends on the sign of the rotation
parameter $\alpha$.  If $\alpha$ is positive, it is
clear that the solution has a naked singularity at $r=0$, cloaked by the
VLS at $r=r_L$ where $B(r_L)$ vanishes.  If instead $\alpha$
is sufficiently negative, the solution will develop a Killing horizon
at $r=r_+< r_L$, inside the VLS.  To avoid a conical singularity at this
pseudo-horizon, it is necessary that the $\td t$ coordinate be periodic,
with period given by
%%%
\be
\Delta \td t = \fft{\pi(g^2 (r_+^4+q)^2 - r_+^6)}{g^3 (3 r_+^8 +
2q r_+^4 - q^2) + 2 g r_+^6}\,.
\ee
%%%%
It is also possible for $V$ to have a double root, in which case the $\td t$
coordinate does not require a periodic identification.  This occurs if
the parameters satisfy the conditions
%%%
\be
g^2=\fft{2r_0^6}{(q+r_0^4)(q-3r_0^4)}\,,\qquad
g\alpha = -\fft{r_0^4(3q-r_0^4)}{2(q-3r_0^4)}\,.
\ee
%%%%
The function $V$ in the metric (\ref{d7equalc}) is then given by
%%%%
\be
V=\fft{(r^2-r_0^2)^2}{g^2 r^2} \Big( 1 + \fft{(q-r_0^4)^2}{2r_0^6 r^2}
+\fft{q^2}{r_0^4 r^4}\Big)\,.
\ee
%%%%
The remaining details of the solution can be obtained by
substituting the $\alpha$ and $g$ parameters given above.  At $r=r_0$,
we have
%%%
\be
B= -\fft{g^2 (q + 5 r_0^4)^2}{16r_0^4}\,,
\ee
%%%
which is negative, implying the occurrence of naked CTC's.

      Let us now consider the case of a critical rotation, such that
the solution contains no naked CTC's.  This can be achieved if $V$ and
$B$ approach zero simultaneously, which occurs if
%%%
\be
\alpha = - \fft{r_0^4}{g}\,,\qquad q=g^{-1} r_0^3 - r_0^4
\,.
\ee
%%%
The metric functions are then given by
%%%
\bea
V &=& (r^2-r_0^2)\Big[g^2 + \fft{1 + g^2 r_0^2}{r^2} -
\fft{r_0^2(g^2 r_0^2 -2 g r_0 -1)}{r^4} -
\fft{r_0^4(g r_0-1)^2}{r^6}\Big]\,,\nn\\
B &=& \fft{r^2-r_0^2}{g^2 r^8 H^2}
\Big[ (r_0^3 + g r_0^2 r^2 + g r^4)^2 -
g^2r_0^2r^2(r^2 + r_0^2)^2\Big]\,,\nn\\
f&=&\fft{(r^2-r_0^2)}{g r^6\, BH^2} 
  \Big[g^2 r^2 (r^2+r_0^2)^2 - r_0^2(gr^2 + g r_0^2 - r_0)^2\Big] \,.
\eea
%%%%
We must then examine the metric in the neighbourhood of
$r= r_0$, where $g_{\psi\psi}\rightarrow 0$, in
order to determine the conditions for regularity.  Defining
$r-r_0=\rho^2$, we find that the metric near $\rho= 0$ becomes
%%%%
\be
ds^2 \sim H(r_0)^{2/5}\Big[ \fft{2 r_0}{1 + 4 g r_0}\Big( d\rho^2 +
(1 + 4 g r_0)^2 \rho^2(\sigma + f\, dt)^2\Big) + \cdots\Big]\,.
\ee
%%%%%
Since $\psi$ has period $2\pi/k$ for $S^5/\Z_k$, it follows that
the quantisation condition
%%%
\be
1 + 4 g r_0= k
\ee
%%%%
must hold.  With this condition, we obtain a completely regular
topological soliton, analogous to the five-dimensional example that
we found section \ref{sec:d5ks}. 

     It is worth remarking that in this case, we have
%%%
\be
H=1 + \fft{q}{r^4} = 1 - \fft{r_0^4}{r^4} + \fft{r_0^3}{g r^4}\,.
\ee
%%%
It follows that there is no naked singularity if $g r_0>0$, even
if $q$ is negative.  This is consistent with the fact that the total
energy $E=\ft18 \pi^2 g^{-1} r_0^3 (g r_0 +4)$ is positive definite
when we have $g r_0>0$.  The thermodynamic quantities for these
topological soliton solutions are given by
%%%%%
\be
E= \fft{\pi^2 r_0^3(4+g r_0)}{8g}\,,\qquad
   J = \fft{\pi^2 r_0^4}{8 g}\,,\qquad
   Q_1=Q_2 = \fft{\pi^2 r_0^3}{4g}\,.
\ee
%%%%%

\subsection{$E + 3g J + \sum Q_i=0$: Case B}

   This condition is satisfied by 
%%%%%
\be
e^{\delta_1 + \delta_2} = 1 - \fft{2}{3ag}\,.
\ee
%%%%%
For simplicity and clarity, we first consider the case with two equal
charges, namely $\delta_1=\delta_2$.  Making the same changes
of the variables and coordinates as in (\ref{d7var1eq}), we find that
the solution becomes
%%%%
\bea
ds^2&=& H^{2/5} \Big(-\fft{V}{H^2 B} r^2 d\td t^2 + B
(\sigma + f d\td t)^2 + \fft{dr^2}{V} + r^2 d\Sigma_2^2\Big)\,,\nn\\
A_\1^1 &=&A_\1^2 = \fft{1}{r^4 H} \fft{q + 2g\alpha}{q-2g\alpha}
(q d\td t +\alpha \sigma)\,,\quad
A_\2 = -\fft{\alpha}{r^4H} d\td t\wedge \sigma\,,\quad
A_\3 = - \fft{\alpha}{r^2}\sigma\wedge J\,,\nn\\
X_1 &=& X_2 = H^{-1/5}\,,\qquad H = 1 + \fft{q}{r^4}\,,\nn\\
V &=& 1 +g^2 r^2 H^2 - \fft{2g\alpha (3q + 2 g\alpha)}{(q-2g\alpha) r^4}
+\fft{8g\alpha^3}{(q-2g\alpha)^2 r^6}\,,\nn\\
B &=& r^2 - \fft{\alpha^2}{r^8 H^2} \Big (1 -
\fft{8 g\alpha}{(q-2g\alpha)^2}r^4\Big)\,,\qquad
f=\fft{1}{B} \Big(g r^2 + \fft{(q+2g\alpha)^2\alpha}{(q-2g\alpha)^2 r^4
H^2}\Big)\,.\label{d7case2}
\eea
%%%
From (\ref{d7res}), the thermodynamic quantities are given by
%%%%
\bea
E&=&\fft{\pi^2(4q^3 -g\alpha q^2 - 4 g^2\alpha^2 q + 4 g^3\alpha^3)}{
8 (q-2g\alpha)^2}\,,\qquad
J= -\fft{\pi^2\alpha (q^2 + 4 g\alpha q -4g^2\alpha^2)}{
8(q-2g\,\alpha)^2}\,,\nn\\
Q_1&=&Q_2=-\fft{\pi^2 (q+2g\alpha)(q-g\alpha)}{4(q-2g\alpha)}
\,.
\eea
%%%%%
In the supersymmetric limit discussed in section \ref{egjqsec}, it was
$K_+=\del/\del \td t$ that had the spinorial square root, corresponding
to having an angular velocity $+g$ at infinity.  By contrast, in
the present case we find that the Killing vector with the spinorial
square root is given by $K_-=\del/\del \td t -2 g \del/\del\psi$, 
corresponding to
having an angular velocity $-g$ at infinity.  We then find
%%%
\bea g(K_-, K_-) = H^{2/5}\Big(-\fft{V}{H^2 B} r^2 + B (f-2g)^2\Big)
=-H^{-8/5} \Big(1 - \fft{2g\alpha}{r^4}\Big)^2\,, \label{11eq}
\eea
%%%%%
which is non-positive, consistent with the supersymmetry.  Thus in general the
horizon, where $V=0$, occurs when $B$ is negative, implying the
occurrence of naked CTC's.

    However, in this case it is possible to arrange that the
right-hand side of equation (\ref{11eq}) is zero at $V=0$.  (This is
not possible for the previous case in (\ref{d7gkk1}).)  Thus
supersymmetric black holes without naked CTC's can arise in the
present case.  The requirement for such a solution can be easily
obtained by requiring that $V(r_0)=0$ and $(1 - 2g\alpha/r_0^4)=0$ 
at some radius $r=r_0$.  This implies that the parameters should
be chosen so that
%%%
\be
\alpha= \fft{r_0^4}{2g}\,,\qquad q^2=r_0^8 + \fft{r_0^6}{g^2}\,.
\label{alqd7}
\ee
%%%%
With these choices, the metric function $V$ is given by
%%%
\be
V = \fft{g^2(r^2-r_0^2)^2}{r^2}\Big(1 + \fft{r_0^8 + q^2}{
r_0^6 r^2} + \fft{r_0^8 + 2 q r_0^4 + 2 q^2}{r_0^4 r^4}
\Big)\,,
\ee
%%%
which, remarkably, in fact has a {\it double} root at $r=r_0$.  At $r=r_0$,
we have that
%%%%
\be
B(r_0) =\ft34 r_0^2 + \ft12 g^2 (q + r_0^4)\,.
\ee
%%%%
It follows from (\ref{alqd7}) that $q>r_0^4$, and so $B(r)$ is
positive on the horizon.  In fact it is straightforward to verify that
$B(r)$ is positive definite at all radii from the horizon at $r=r_0$
to $r=\infty$.  Thus the metric (\ref{d7case2}) with the parameters
satisfying (\ref{alqd7}) describe a supersymmetric black hole that is
regular everywhere on and outside the horizon, with the near-horizon
geometry being the product of AdS$_2$ and a squashed $S^5$.  The
solution preserves $\ft18$ of the supersymmetry.  The double root of
$V(r)$ at $r=r_0$ implies that the supersymmetric black hole has zero
temperature.  This seven-dimensional supersymmetric black hole is
analogous to the five-dimensional one found in \cite{gutrea1}, which
we discussed in section \ref{e2gjqsec}.

    As in the five-dimensional case discussed in section
\ref{e2gjqsec}, an alternative way to avoid CTC's is to consider the
possibility that both $V$ and $B$ vanish at some radius $r=r_0$.
This can be achieved by choosing the parameters so that
%%%%
\bea
q&=&\fft{g\alpha^3 (1 + 4 g^2 r_0^2) + \alpha^2 r_0^4 (1 - 3 g^2 r_0^2)
+g\alpha r_0^{10}}{r_0^2 (3g^2 \alpha^2 - 4 g\alpha r_0^4 + r_0^8}
\,,\nn\\
0&=&g^2\alpha^4(1-2g^2r_0^2)^2 +2 g\alpha^3 r_0^4(1 + 15g^2 r_0^2 +
2 g^4 r_0^4) - \alpha^2 r_0^8 (-1 + 24 g^2 r_0^2 + 3 g^4 r_0^4)
\nn\\
&&+2 g\alpha r_0^{14} (3 - g^2 r_0^2) + g^2 r_0^{20}\,.
\eea
%%%%
Thus, for the cases where $r_0$ is the largest root for both
$V$ and $B$, the solution describes a smooth
supersymmetric soliton.

   The remaining solutions inevitably have naked CTC's.  For those
with a Killing horizon, {\it i.e.} $V=0$ at some $r=r_0 >0$ outside
the singularity at $r=0$, the $\td t$ coordinate must be appropriately
periodically identified, in order to avoid a conical singularity at
$r=r_0$.  Having done so, the geodesics are then complete from $r=r_0$
to $r=\infty$.  As usual, if $V$ has a {\it double} root at $r=r_0$, then
the $\td t$ coordinate does not require periodic identification.  A double
root is achieved by choosing the parameters to be given by
%%%
\be
\alpha = -\fft{(16 g^2 r_0^2 + 9)(4g^2 r_0^2 +3)^2}{32 g^5}\,,\qquad
q=\fft{3(4g^2 r_0^2)^2}{16 g^4}\,.
\ee
%%%%
The metric function $V$ becomes
%%%
\be
V=\fft{g^2(r^2-r_0^2)^2}{r^2} \Big(1 + \fft{2g^2 r_0^2}{g^2 r^2} +
\fft{72 g^4 r_0^4 + 88 g^2 r_0^2 + 27}{8g^2 r^4}\Big)
\ee
%%%
It is straightforward to verify that at $r=r_0$, the function $B(r)$
is negative, implying naked CTC's.  Thus the solution describes
a supersymmetric BMPV type of repulson.

       For non-equal charges, the solution becomes much complex, but
the structure is very similar.  The metric can be cast into the form
%%%
\be
ds^2 = (H_1 H_2)^{\ft15} \Big( -\fft{V}{H_1 H_2 B} r^2 d\td t^2+
B(\sigma + f d\td t)^2 + \fft{dr^2}{V} + r^2 d\Sigma_2^2\Big)\,,
\ee
%%%
where $\td t=t/\Xi_-$ and $r=R/\sqrt{\Xi}$.  The simplest way to determine
if there exists a supersymmetric black hole is to examine the norm
of the Killing vector $K_-=\del/\del_{\td t} - 2g\, \del/\del_\psi$;
it is given by
%%%
\bea
g(K_-, K_-) &=& (H_1H_2)^{\ft15} \Big(-\fft{Y}{f_1} + B^2 (f-2g)^2\Big)
\,,\nn\\
&=& -(H_1H_2)^{-4/5}\Big(1 - \fft{162(e^{2\delta_1}-1)(
e^{2\delta_2}-1) (e^{\delta_1 +\delta_2} -1)^4}{
e^{\delta_1 + \delta_2} (3 e^{\delta_1 + \delta_2} -5)^2
(3 e^{\delta_1 + \delta_2} -1)^3\, r^4}\Big)\,,\label{d7gkk}
\eea
%%%%
which is non-positive.  Thus naked CTC's can only be avoided in  two
circumstances.  One is that the Killing horizon is coincident with
the VLS,  \ie $V=0=B$ at some radius  $r=r_0$.  This
leads to a regular supersymmetric soliton.

     Alternatively, we can require that at $V=0$, the right-hand side
of (\ref{d7gkk}) also vanishes.  This can be achieved if we take
%%%
\be
m=\fft{128e^{\delta_1 + \delta_2} (3e^{\delta_1 + \delta_2}-1)^2}{
729 g^4 (e^{2\delta_1}-1)(e^{2\delta_2}-1) (e^{\delta_1 + \delta_2}+1)^2
(e^{\delta_1 + \delta_2} -1)^4}\,,
\ee
%%%%
Then remarkably, the function $V$ has a double root at $r=r_0$, given
by
%%%
\be
r_0^2 = \fft{16}{3 (e^{\delta_1+\delta_2} +1)(3e^{\delta_1 + \delta_2}
-5)g^2}\,.
\ee
%%%
With these choices of charge parameters, we find that the function
$V$ now becomes
%%%
\be
V= \fft{g^2(r^2-r_0^2)^2}{r^2}\Big( 1 + \fft{9 
e^{2(\delta_1 +\delta_2)} - 6 e^{\delta_1 + \delta_2} + 17}{
3(e^{\delta_1 + \delta_2} +1)(3e^{\delta_1 +\delta_2}-5) g^2 r^2}
+ \fft{h}{g^4 r^4}\Big)\,,
\ee
%%%
where $h$ is a constant, given by
%%%
\bea
h&=&\Big[32(-2d_1^2 - 2d_2^2 + 9 d_1 d_2 + 9 d_1^5 d_2^5 -3d_1^3 d_2^3(
d_1 + d_2)^2 + 2 d_1^2d_2^2(2d_1^2 - 3 d_1 d_2 + 2d_2^2)\nn\\
&& -d_1 d_2 (3 d_1^2 - 2 d_1 d_2 + 3 d_2^2))\Big]/
\Big[9d_1d_2(d_1^2-1)(d_2^2-1)(d_1d_2+1)(3d_1d_2-5)^2\Big]\,,
\eea
%%%
with $d_1=e^{\delta_1}$ and $d_2=e^{\delta_2}$.
The function $B(r)$ at $r_0$ is positive, and it is positive for all
$r\ge r_0$, implying a supersymmetric black hole regular on and outside
the horizon.  This is the unequal-charge generalisation of the equal-charge
regular black holes that we obtained above.  These seven-dimensional 
supersymmetric black holes are analogous to the five-dimensional ones
obtained in \cite{gutrea2}, which we discussed in section \ref{e2gjqsec}. 
Note that the right-hand side of (\ref{d7gkk}) is
negative definite if we turn off either of the charges, in which case
there can be no supersymmetric black holes.

\section{Rotating Black Holes and Supersymmetric Limits in 
Four-Dimensional Gauged Supergravity}

\subsection{Black-Hole Thermodynamics in Four Dimensions}

    Charged rotating black holes in four-dimensional Einstein-Maxwell
theory with a cosmological constant were found in \cite{carter3}.  
Recently, generalisations were obtained which can be viewed as charged rotating
black holes in four-dimensional ${\cal N}=4$ gauged supergravity, with
independent charges carried by the two gauge fields in the $U(1)\times
U(1)$ abelian subgroup of the $SO(4)$ gauge group \cite{d4gauge}.  They can
also, therefore, be viewed as solutions in ${\cal N}=8$ gauged supergravity,
where the four charges associated with the $U(1)^4$ abelian subgroup of
$SO(8)$ are set pairwise equal.  The truncation of the ${\cal N}
=4$ Lagrangian to the relevant sector for describing these solutions
is given by
%%%%%
%%%%%
\bea
{\cal L}_4 &=& R\, {*\oneone} -\ft12 {*d\varphi}\wedge d\varphi -
\ft12 e^{2\varphi}\, {*d\chi}\wedge d\chi - \ft12 e^{-\varphi}\,
 {*F_{\2 2}}\wedge F_{\2 2} -\ft12 \chi\, F_{\2 2}\wedge F_{\2 2} \nn\\
&& -\fft1{2(1+\chi^2\, e^{2\varphi})}\, (
   e^{\varphi}\, {*F_\2 1}\wedge F_{\2 1} - e^{2\varphi}\, \chi\,
       F_{\2 1}\wedge F_{\2 1})\nn\\
&&
-g^2\, (4 + 2 \cosh\varphi + e^{\varphi}\, \chi^2)\, {*\oneone}
\,.\label{d4lag}
\eea
%%%%%
The non-extremal rotating charged black hole solutions are given, 
in a frame that rotates at infinity, by 
\cite{d4gauge}
%%%%%
\bea
ds_4^2\!\!\!\ &=&\!\!\! -\fft{\Delta_r}{W}\, 
(dt - a\, \sin^2\theta\,\Xi^{-1} 
d\phi)^2 + W\,  \Big( \fft{dr^2}{\Delta_r} +
\fft{d\theta^2}{\Delta_\theta} \Big)
   + \fft{\Delta_\theta\, \sin^2\theta}{W} [a dt - (r_1 r_2 +
   a^2) \Xi^{-1} d\phi]^2\,,\nn\\
e^{\varphi_1}\!\!\! &=&\!\!\! \fft{r_1^2 + a^2\cos^2\theta}{W}=
1 + \fft{r_1\, (r_1-r_2)}{W}\,,\qquad
   \chi_1 = \fft{a\, (r_2-r_1)\, \cos\theta}{r_1^2 + a^2\,
     \cos^2\theta}\,,\nn\\
A_{\1 1} &=& \fft{2\sqrt2 m\, s_1\, c_1\, r_2\, (dt-a\, \sin^2\theta\,
  \Xi^{-1}\, d\phi)}{W}\,,\nn\\
A_{\1 2} &=& \fft{2\sqrt2 m\, s_2\, c_2\, r_1\, (dt-a\, \sin^2\theta\,
  \Xi^{-1}\, d\phi)}{W}\,,\label{d4sol}
\eea
%%%%%
where
%%%%%
\bea
r_i &=& r + 2m\, s_i^2\,,\nn\\
\Delta_r &\equiv& \Delta + g^2 \, r_1\, r_2\, (r_1\, r_2 + a^2)
=r^2 + a^2 - 2m\, r + g^2 \, r_1\, r_2\, (r_1\, r_2 +
a^2)\,,\nn\\
\Delta_\theta &\equiv& 1 - g^2\, a^2\, \cos^2\theta\,,\qquad
W=r_1\, r_2 + a^2 \cos^2\theta\,,\qquad \Xi=1-a^2\, g^2\,.
\eea
%%%%%
As usual, $s_i=\sinh\delta_i$ and $c_i=\cosh\delta_i$.
Note that we are using the ``undualised'' form of the four-dimensional 
theory here, where, as discussed in \cite{d4gauge}, all the charges
are electric.  Note also that we have rescaled the azimuthal coordinate 
$\phi$ by a factor of $\Xi^{-1}$ here, relative to the normalisation
used in \cite{d4gauge}, so that $\phi$ in (\ref{d4sol}) has the
canonical period $2\pi$.

   In order to give a more uniform treatment of these four-dimensional
solutions that harmonises with our discussion in five and seven dimensions, 
we shall adopt a maximal supergravity notation at this point,
and view the solution (\ref{d4sol}) as a 4-charge solution with pairwise
equal charges.  This will avoid the necessity for $\sqrt2$ factors 
associated with the charges.  Thus we shall have charges $Q_1=Q_2$ 
characterised by the parameter $\delta_1$, and charges $Q_3=Q_4$ 
characterised by $\delta_2$.  This change of viewpoint will be understood
in all our subsequent formulae for charges and electrostatic potentials.

  The coordinates in (\ref{d4sol}) are rotating at infinity.  A non-rotating
coordinate system is achieved by defining a new azimuthal angle 
$\phi' = \phi + a\, g^2\, t$.  The time coordinate $t$ has the canonical
normalisation.  It is helpful to recast the metric (\ref{d4sol}) in the form
%%%%%
\be
ds_4^2 = -\fft{\Delta_r\, \Delta_\theta}{B\Xi^2}\, dt^2 
  + B\, \sin^2\theta\, (d\phi + f\, dt)^2 
   + W\, \Big(\fft{dr^2}{\Delta_r} + \fft{d\theta^2}{\Delta_\theta}\Big)\,,
\ee
%%%%%
The entropy, temperature, angular velocity and 
electrostatic potentials on the horizon are given by
%%%%%
\bea
S&=& \fft{\pi(r_1 r_2 + a^2)}{\Xi}\,,\qquad
T = \fft{\Delta_r'}{4\pi (r_1 r_2 + a^2)} \,,\qquad
\Omega= \fft{a(1+g^2\, r_1 r_2)}{r_1 r_2 + a^2}\,,\nn\\
\Phi_1 &=& \Phi_2 =\fft{2ms_1 c_1\, r_2}{r_1 r_2 + a^2}\,,\nn\\
\Phi_3 &=& \Phi_4 =\fft{2ms_2 c_2\, r_1}{r_1 r_2 + a^2}\,,
\eea
%%%%%
where all quantities are evaluated on the outer horizon at $r=r_+$,
the largest root of $\Delta_r$.
Calculating the angular momentum, and the charges, as surface
integrals at infinity, we can then integrate the first law
%%%%%
\be
dE = T\, dS + \Omega\, dJ + \sum_i \Phi_i\, dQ_i\,,
\ee
%%%%%
to obtain the energy.  Our results are
%%%%%
\bea
E &=& \fft{m}{\Xi^2}\, (1+ s_1^2 + s_2^2) = 
  \fft{m}{2\Xi^2}\, (\cosh2\delta_1 + \cosh2\delta_2)\,,\nn\\
J &=& \fft{m\, a}{\Xi^2}\, (1+s_1^2 + s_2^2)= \fft{ma}{2\Xi^2}\, 
              (\cosh2\delta_1 + \cosh2\delta_2)\,,\nn\\
Q_1 =Q_2&=& \fft{m\, s_1\, c_1}{2\Xi}= \fft{m}{4\Xi}\, \sinh2\delta_1\,,\nn\\
Q_3=Q_4 &=& \fft{m\, s_2\, c_2}{2\Xi}= \fft{m}{4\Xi}\, 
             \sinh2\delta_2\,.\label{d4ejq}
\eea
%%%%% 

\subsection{Supersymmetric Limits and the Supersymmetric Bound}

   The algebra of the supercharges ${\cal Q}$ in four-dimensional 
AdS supergravity is given by
%%%%%
\be
M\equiv 
\{ {\cal Q}, \overline{{\cal Q}} \} = \ft12 J_{AB}\, 
        \gamma^{AB} 
             + Z\,,
\ee
%%%%%
where $J_{04}$ is the energy, and $J_{ij}$ for $1\le i\le 3$ correspond 
to angular momenta. Taking $J_{12}=J$ non-zero, we shall have
%%%%%
\be
J_{04}= g\, E\,,\qquad J_{12} = g^2\, J\,,
  \qquad Z =  g\, \sum_i Q_i\,,
\ee
%%%%%
after using the adjoint action of the AdS and R-charge symmetries to
choose a convenient frame for $Z^{IJ}$.
The four eigenvalues of the \bog matrix $g^{-1}\, M_\alpha{}^\beta$ are 
given by
%%%%%
\be
\lambda = E \pm g\, J  \pm \sum_i Q_i\,.\label{d4bog}
\ee
%%%%%
The four eigenvalues are equivalent, modulo sign reversals of the angular
momentum and the charges, so unlike the $D=5$ and $D=7$ cases discussed
previously, there is only one inequivalent case of interest to consider here. 

   The general analysis of the convex cone implied by the positivity
of the \bog matrix eigenvalues described in section \ref{d5bogsec}
applies to this $D=4$ case as well. There is just one angular momentum
and so the cone lies in $\R^3$ and is bounded by two planes.

   Substituting our results (\ref{d4ejq}) for the energy, angular
momentum and charges of the rotating black holes (\ref{d4sol}) 
into (\ref{d4bog}), we find that a zero eigenvalue is achieved if
%%%%%
\be
e^{2\delta_1 + 2\delta_2} = 1 + \fft{1}{2ag}\,.
\ee
%%%%%
The solution preserves $\ft14$ of the supersymmetry.

       To discuss the global structure of the solution, we
can examine the metric function $\Delta_r$, which after imposing
the supersymmetry condition, can be expressed
as a sum of two squares:
%%%
\bea
\Delta_r &=& g^2 \Big(r^2 + m (\sinh^2\delta_1 + \sinh^2\delta_2) \, r +
4 m^2 \sinh^2\delta_1\, \sinh^2\delta_2 +g^{-2}
(\coth(\delta_1 + \delta_2) -1) \Big)^2\nn\\
&& + \tanh^2(\delta_1 + \delta_2)\Big(r  -
\fft{2m\sinh\delta_1\,\sinh\delta_2}{\cosh(\delta_1+\delta_2)}
\Big)^2\,.
\eea
%%%
Thus in general the function $\Delta_r$ has no root, and hence
the solution has a naked singularity.  The only possible root
is given by
%%%
\be
r_+= \fft{2m \sinh\delta_1\,\sinh\delta_2}{\cosh(\delta_1+
\delta_2)}\,,
\ee
%%%
which is achieved by taking
%%%
\be
m\,g=\fft{\cosh (\delta_1 + \delta_2)}{e^{\ft12(\delta_2 +\delta_2)}
\sinh^2(\delta_1 + \delta_2) \sinh (2\delta_1) \sinh(2\delta_2)}
\,.
\ee
%%%
The function $\Delta_r$ then has a double root $r_+$, and hence the
solution describe a supersymmetric black hole that is regular on and
outside the horizon, with zero temperature.  In the case where the
charge parameters are set equal, this solution reduces to the regular
supersymmetric AdS black hole found in \cite{kosper}.

\section{G\"odel  Black Holes}

   In this section we shall apply some of the ideas of this paper to a
closely related set of solutions of ungauged supergravity that are of
some current interest, namely G\"odel black holes
\cite{god1,god2,god3,god4,god5}.  Although the solutions themselves
are not new and have received some considerable discussion, the
particular point we are making appears to have been missed in the
literature.

    The G\"odel black hole solutions of the
ungauged 5-dimensional supergravity theory are given by
%%%%%
\bea
ds^2 &=& -\fft{\Delta}{r^4\, \beta}\, dt^2 + \ft14 \beta r^2 \Big(\sigma_3 -
\fft{4(a m + b r^4)}{r^4 \beta}\, dt\Big)^2 + \fft{r^4 dr^2}{\Delta} 
   + \ft14 r^2 (\sigma_1^2+\sigma_2^2)\,,\nn\\
A &=& \fft{\sqrt3}{2}\, b r^2 \sigma_3\,, \label{adsgod}
\eea
%%%%%
with
%%%%%
\ben
\Delta = r^4 - (2m - 16 b^2 m^2 -8 mab )r^2 +2m a^2\,,\qquad
\beta = 1- 8m b^2 + \fft{2m a^2}{r^4} - 4 b^2 r^2\,.
\een
%%%%%
If $m$ is set to zero, we obtain the pure G\"odel background, whose spatial
sections are flat. We shall discuss this metric in more detailo in 
section \ref{heissec}.

   Expressions for the the energy $E$ , angular momentum $J$,
and charge $Q$ have been obtained in \cite{BarnichCompere}:
%%%%%
\bea
E&=&{ 3 \pi \over 4} m-8 \pi b^2 m^2 -\pi abm\,,\\
J&=& \ft12 \pi ma - \pi m a^2 b - 4 \pi b^2 m^2 a\,\\
Q&=& 2 \sqrt{3} \,\pi m ab\,.
\eea
%%%%%
The solution is singular at $r=0$; both the curvature and the field strength
diverge there.  The latter can be seen from
%%%%%
\be
F^2 = \fft{48 b^2 (r^2-m)}{r^2}\,.
\ee
%%%%%
Note that the gauge field is magnetic at large distances, but becomes
electric for $r^2<m$.

   Since $g_{tt}= 2m/r^2- 1$, there is an ergo-region when
$r^2<2m$.  Since 
%%%%%
\ben 
g_{\psi \psi} ={ r^2 \over 4} \beta(r)\,,
\een 
%%%%%
there is a VLS, situated at $r=r_L$, where 
%%%%%
\ben 
\beta(r_L)=0\,.  
\een
%%%%%

    If $a=0$,  the VLS is located at
$r^2 =r_L^2= (1-8mb^2)/(4 b^2)$.  There are no  CTC's
within  the domain $r< r_L$, and so the time machine occupies the region
$r>r_L$. The Killing horizon
is situated at  $r^2=r_H^2= 2m (1-8m b^2)$,  which is always
inside the ergo-region. One has
%%%%%
\ben
r_L^2-r_H^2 = {(1- 8mb^2 )^2 \over 4b^2 }\,.
\een
%%%%%
 
    By contrast, if  $a\ne 0$ the situation is more complicated.
The horizon is now situated at
%%%%%
\ben
r_H^2= m-4 mab - 8b^2 m^2 + m \sqrt \alpha\,,
\een
%%%%%
with
%%%%%
\ben
\alpha = (1-8b^2m ) (1-8b^2 m-8ab - { 2a^2 \over m} ) \,.
\een
%%%%%
One has entropy, angular velocity and surface gravity given by
%%%%%
\bea
S &=& \ft12 \pi ^2 r_H^3 \sqrt {\beta(r_H)}\,,\nn\\
\Omega &=&  { 4br_H^2  + 4m a  \over r_H^2\, \beta (r_H) }\,,\nn\\
\kappa &=& { 2m \sqrt \alpha \over r_H^3 \sqrt{\beta(r_H)} }\,.
\eea
It was verified in  \cite{BarnichCompere} that
the first law of thermodynamics holds for these expressions.
Thus the usual thermodynamic interpretation
holds, as long as the black hole lies in inside the region
where there are no CTC's.

   One may instead consider the case where the horizon moves into the time 
machine, \ie  $\beta(r_H)<0$ and thus $r_H>r_L$.  In this
case, the horizon is really to be thought of as a pseudo-horizon
that closes off the spacetime.  Thus the radial coordinate cannot
exceed $r_H$, and at $r=0$ there is a naked singularity.
The surface gravity is now purely imaginary, and so
the (real) time $t$ must be identified with the real period
%%%%%
\ben
\Delta t= { 2 \pi r_H^3 \sqrt{|\beta(r_H)|}  \over 2m \sqrt \alpha }\,.
\label{tper}
\een
%%%%%

   A point apparently missed in the literature on G\"odel black holes
is that the gauge field given in (\ref{adsgod}) 
becomes singular at $r=r_H$, which may be seen from the fact that
%%%%%
\be
A_\mu\, A_\nu\, g^{\mu\nu} = \fft{3 b^2 r^4(r^2-2m)}{\Delta}\,.
\ee
%%%%%
If $r_H<r_L$ we may pass to a new gauge potential
%%%%%%
\be
A'= A - \fft{2\sqrt3 b (am+b r_H^4)}{r_H^2\, \beta(r_H)}\, dt\,,
\ee
%%%%%
for which $|A'|^2 = P(r)/\Delta(r)$, 
where $P(r)$ is a poynomial of degree three in $r^2$ 
which vanishes at $r=r_H$. If $r_H<r_L$ and 
if $r_H$ is the largest positive root of 
$\Delta$, then the gauge potential $A'$ will be regular everywhere
outside and on the horizon.  If $r_H>r_L$, and the next-smallest root 
$r_-$ of 
$\Delta$ is smaller than $r_L$, the gauge potential is bounded for 
$r_- < r \le r_H$. However, because we now have a pseudo-horizon, for
which the coordinate $t$ must be identified with period given by (\ref{tper}),
we obtain a Josephson quantisation condition of the form 
%%%%%
\be
\fft{\sqrt3\, e b (am+ b r_H^4) r_H}{m\sqrt{\alpha\, \beta(r_H)}}\,,
\ee
%%%%%
if there are fields of charge $e$ present.

   In ungauged supergravity, the Dirac or Josephson quantisation
conditions need not hold, since the fields in the supergravity
multiplet are uncharged.  There does also exist a five-dimensional 
G\"odel-AdS type solution in gauged supergravity, which was found in 
\cite{behpos} and is given by
%%%%%
\bea
ds ^2 &=& -(dt + \omega)^2 + 
  {dr^2 \over 1+g^2 r^2} + {r^2 \over 4} \Bigl( \sigma_1^2 + \sigma_2^2 +
  (1+ g^2 r^2)\sigma_3^2 \Bigr )\,,\nn\\
\omega &=& {r^2 \over 2} \bigl(g \sigma_3 - {h \over 1+g^2 r^2} 
                    \sigma_1\bigr)\,,\nn\\
A &=& { {\sqrt 3} h r^2 \over 2 (1+g^2 r^2 ) }\sigma _1\,.\label{bpos}
\eea
%%%%%
The field strength has norm given by
%%%%%
\be
F_{\mu\nu}\, F^{\mu\nu} = \fft{48 h^2}{(1+g^2 r^2)^3}\,.
\ee
%%%%%   
If $h=0$, we get the manifestly $U(2) \times {\Bbb R}$ invariant
AdS$_5$ metric with a Bergmann base.
If $h\ne 0$ the manifest $U(2)$ is broken to $SU(2)$.  If instead $g$ is
set to zero, we obtain the pure G\"odel background, which will be discussed
in section \ref{heissec}.

    The metric induced on the $SU(2)$ orbits, which are  squashed at 
infinity,  is 
%%%%%
\be
{ r ^ 2\over 4} \Bigl (   \sigma_2 ^2 + \bigl (\sigma_3  -{gh r^2 \over
1+g^2 r^2 } \,  \sigma _1 \bigr )^2 + {1+ (g^2 -h^2) r^2 \over 1+g^2
r^2 } \, \sigma _1^2  \Bigr )\,.
\ee
%%%%%
Thus the orbits of $SU(2)$ remain spacelike at large $r$, or become
timelike, depending on whether
%%%%%
\be
g^2 > h^2\,, \qquad {\rm or } \qquad  h^2> g^2\,
\ee
%%%%%
respectively.

    In both cases the coordinate $t$ is globally defined, and there
is no need to make an identification.  In fact
%%%%%
\be
g^{tt} = g^{\mu\nu}\del_\mu t\, \del_\nu t=
 -\fft{[1+(g^2-h^2) r^2]}{(1+g^2 r^2)^2}\,.\label{tfun}
\ee
%%%%%
In the first case, $\del_\mu t$ is always timelike, and therefore 
the coordinate $t$ is a global time function, \ie it increases
along every future-directed timelike curve, and 
there are no CTC's. In the latter case, $\del_\mu t$ ceases to be 
timelike outside a VLS, which is located at 
%%%%%
\be
r^2 = {1 \over h^2-g^2}\,.
\ee
%%%%%
Thus in this case, while being a globally-defined coordinate, $t$ is
not a global time function. 
The norm of the potential $A$ given in (\ref{bpos}) is
%%%%%
\be
A_\mu\, A_\nu\, g^{\mu\nu} = \fft{3 h^2 r^2}{(1+g^2 r^2)^2}\,,
\ee
%%%%%
which is regular everywhere.  We conclude that since $A$ is globally
defined, there is no need to impose any quantisation condition.  One might
wonder whether, if one chose to identify $t$ periodically, a quantisation
condition would result.  However, because the potential $A$ is 
globally well-defined even if $t$ is periodically identified, no 
quantisation condition would arise in this case either.  Curiously, 
there exist two globally well-defined gauges in which the new gauge
potentials $A'$ fall off faster at infinity than does $A$, which falls
of like $1/r^2$.  Namely, if we define
%%%%%
\be
A' = A - \fft{\sqrt3\, h}{h \pm g}\, dt\,,
\ee
%%%%%
then we find
%%%%%
\be
A'_\mu\, A'_\nu\, g^{\mu\nu} = -\fft{3 h^2}{(h\pm g)^2\, (1+g^2 r^2)^2}\,.
\ee
%%%%%
Since both of these gauges are globally well-defined, no quantisation 
condition would arise, even if we insisted upon the faster fall-off
that they exhibit.

    This analysis of transition functions is supported by the observation
that the topology of the solution (\ref{bpos}) is trivial; it is the
product, topologically, of the time $\R$ and the 
Bergmann manifold, which itself is topologically $\R^4$.

\subsection{Heisenberg Quantization Conditions}\label{heissec}

   While on the subject of quantisation conditions, 
it may be of interest to reconsider the pure
 G\"odel solution. This is the ground state with respect to which
the energy, angular momentum and charge of the G\"odel black holes 
are measured. Since one must pass through the VLS in order to travel backwards 
in time, one might consider compactifying the spatial sections so that
a unit cell lies inside the VLS, in order to prevent time travel.
It turns out that one may indeed compactify the spatial sections, but
only at the expense of passing to a periodic time coordinate.

  To see this, note that the pure G\"odel ground state metric may be 
cast in the form
%%%%%
\ben
ds ^2= -\bigl
[dt + {2 b } (xdy-ydx + z dw - wdz) \bigr]^2 + 
      dx ^2 + dy ^2 + dz^2 + dw^2\,.
\een
%%%%%
The metric is homogeneous, since it admits the five Killing fields
 $ R_t =\p_t$ and
%%%%%
\ben
R_x = \p_x - { 2b}\, y \p_t\,\qquad R_y =  \p_y + {2b }\,x\, 
 \p_t \,\qquad R_z =
 \p_z - {2b}\,  w \p_t\,\qquad R_w =
 \p_w + { 2b}\,  z\p_t
\een
%%%%%

One checks that time translation is central,
and moreover, the only non-vanishing  brackets are
%%%%%
\ben
[R_x,R_y]= {4b}  R_t\,\qquad [R_z,R_w]= {4b }  R_t \,.
\een
%%%%%
This gives a Heisenberg type algebra $\frak{g}$.
In fact one may regard the metric as a left-invariant metric on
the Heisenberg type group $G$.

The one forms $ dt + {2b } (xy-ydx + z dw - wdz)$ and $dx,dy,dz,dw$
are left-invariant. The vector fields $R_t,R_x,R_y,R_z,R_w$ generate
left translations, and are right-invariant.

   The gauge field supporting the metric, 
%%%%%
\ben
A= {\sqrt{3} \over 2} \bigl(dt + {2 b } (xdy-ydx + z dw - wdz) \bigr)
 -  {\sqrt{3} \over 2} dt\,,\label{godela1}
\een
%%%%%
is invariant only up to a time-dependent gauge transformation because
%%%%%
\ben
\pounds_{R_x} dt = -dy\,, \qquad \pounds_{R_y } dt = dx\,.
\een
%%%%%
The gauge potential (\ref{godela1}) satisfies
%%%%%
\be
A_\mu\, A_\nu\, g^{\mu\nu} = 3b^2\, (x^2 + y^2 + z^2 + w^2)\,.
\ee
%%%%%
We note that the norm of the field strength is given by
%%%%%
\be
F_{\mu\nu}\, F^{\mu\nu} = 48 b^2\,.
\ee
%%%%%

   One could pass to  a new gauge, in which the transformed gauge potential
$\td A$ is invariant, by making the gauge transformation
%%%%%
\ben
A \rightarrow \tilde A=  A + d \bigl ( {\sqrt{3} \over 2} t \bigr )\,.
\een
%%%%%
The new gauge field $\tilde A $ has constant magnitude
%%%%%
\ben
\tilde A_\mu \, \tilde A_\nu\, g^{\mu\nu} = - {3 \over 4}\,.
\een
%%%%%
Acting on a field $\Psi$ of charge $e$, the necessary gauge transformation
is
%%%%%
\ben
\Psi \rightarrow e^{(i e  {\sqrt{3} \over 2}) t} \Psi\,.
\een
%%%%%

    In order to discuss the identifications
it is helpful  to introduce some convenient notation.
If $V$ is a vector field, the operator $e^{\lambda V}$
acting on functions  gives
%%%%%
\ben
e^{\lambda V }f(x ^\mu ) = f( \tilde x ^\mu )\,,
\een
%%%%%
where $\tilde x$ is the point obtained by moving a parameter distance
$\lambda $ along the integral curves of $V$.
In other words $\tilde x^\mu = x ^\mu (\lambda)$, where
%%%%%
\ben
{dx ^\mu  \over d \lambda }= V ^\mu (x) \,,\qquad x^\mu(0)= x^\mu\,.
\een
%%%%%
We could write
%%%%%
\ben
\tilde x ^\mu = e^ {\lambda V} x ^\mu \,.
\een
%%%%%
Thus, for example,
\bea
e^{\lambda R_x}f(t,x,y,z,w)& =&
 f(t- {2b y \lambda  },x+\lambda,y,z,w)\,,\\
 e^{\lambda R_y}f(t,x,y,z,w) &=& f(t+{2 b x \lambda },x,y+ \lambda ,z,w)\,.
\eea
%%%%%
If $\phi_\lambda$ is the one-parameter group of diffeomorphisms
associated to the vector field $V$,  the usual definition of pull-back
becomes in this notation
%%%%%
\ben
\phi_\lambda ^\star f(x) = f(\phi ^{-1} _\lambda (x))= e^{-\lambda V} f(x)\,.
\een
%%%%%

   Now consider attempting to identify the $x$ coordinate,
with period $d_1 $ say.   As it stands, this is not a symmetry, and therefore
one must shift in time as well, \ie we   demand that
%%%%%
\ben
(t,x,y,z,w) \equiv (t-2by d_1 , x+d_1, y, z,w) \,.
\een
%%%%%
If we instead identify the $y$ coordinate,
with period $d_2$, we must demand that
%%%%%
\ben
(t,x,y,z,w) \equiv (t+2bx d_2 , x,y +d_2,  z,w) \,.
\een
%%%%%
However, these two identifications do not commute.
In fact, one easily checks from the Lie algebra that
%%%%%
\ben
e^{d_1 R_x} e^{d_2 R_y }  e ^{-d_1 R_x} e^{-d_2 R _y} =
e ^ {4 d_1 d_2 b R_t}\,.
\een
%%%%%
One must therefore identify the time coordinate as well, \ie demand
that
%%%%%
\ben
(t,x,y,z,w) \equiv (t+4b d_1d_2 , x,y,  z,w) \,.
\een
%%%%%
Similar considerations apply if one wishes to identify $z$ and $w$,
with periods $d_3$ and  $d_4 $ say. This also entails an identification
of $t$, with  period ${4b d_3d_4}$. Consistency then  requires that
%%%%%
\ben
d_1d_2  =d_3 d_4  \,. \label{compact}
\een
%%%%%

\section{Conclusions}

    In this paper, we have studied the thermodynamics of the
recently-discovered non-extremal charged rotating black holes of
gauged supergravities in five, seven and four dimensions, obtaining
energies, angular momenta and charges that are consistent with the
first law of thermodynamics.  We studied their supersymmetric limits, by
using these expressions together with a Bogomolny analysis of the AdS
superalgebras.  We gave a general discussion of the global structure
of such solutions, and applied it in the various cases.  We obtained new
regular supersymmetric black holes in seven and four dimensions, as
well as reproducing known examples in five and four dimensions.  We
also obtained new supersymmetric non-singular topological solitons in
five and seven dimensions. The rest of the supersymmetric solutions
either have naked singularities or naked time machines. The latter can
be rendered non-singular if the asymptotic time is periodic.  This
leads to a new type of quantum consistency condition, which we call a
{\it Josephson quantisation condition}. Finally, we discussed some
aspects of rotating black holes in G\"odel universe backgrounds.

\section*{Acknowledgement}
   
   We are grateful to Ergin Sezgin for discussions.
G.W.G. thanks the George P. \& Cynthia W. Mitchell Institute for
Fundamental Physics, and H.L. and C.N.P. thank the Physics Department
at the University of Pennsylvania, for hospitality during the course
of this work. We are grateful to G. Compere, G. Barnich, A. Gomberov
and M. Banados for pointing out various misprints and some errors in 
an earlier version of this paper.  During the completion of these revisions,
we were sent a preliminary draft of a paper by Simon Ross discussing
the regularity of some of the metrics in this paper.  Some of the
comments in the draft about our paper no longer apply, in the light of
our revisions.

\appendix
\section{The Spin$^c$ Structure of the Taub-BOLT Manifold}
   
    In this appendix, we give a brief discussion of the spin$^c$ structure
of the $k=1$ $\R^2$ bundle over $S^2$, which is the manifold of the
Taub-BOLT instanton found in \cite{pagebolt}.  This serves as a useful 
illustrative example that exhibits some of the same essential features 
that arise in the topological soliton solutions of section 3.3.1 in the
case that the integer $k$ is odd.

    The Ricci-flat Taub-BOLT instanton metric is given by \cite{pagebolt}
%%%%%
\be
ds^2 = \fft{(r^2-\ell^2)\, dr^2}{(r-2\ell)(r-\ft12\ell)} +
  \fft{4\ell^2\, (r-2\ell)(r-\ft12\ell)}{r^2-\ell^2}\, \sigma_3^2
   + (r^2-\ell^2)\, (\sigma_1^2 +\sigma_2^2)\,,\label{tbmet}
\ee
%%%%%
where $r\ge 2\ell$, and the Euler angles in the $SU(2)$ left-invariant 1-forms
have the standard periods, $0\le\phi< 2\pi$, $0\le\psi <4\pi$, and
$0\le\theta\le\pi$.  There are two $L^2$ harmonic forms, which are 
self-dual and anti-self-dual respectively, and given locally
by $F_i=dA_i$, where the potentials are
%%%%%
\be
A_1 = \Big(\fft{r+\ell}{r-\ell}\Big)\, \sigma_3\,,\qquad
A_2 = \Big(\fft{r-\ell}{r+\ell}\Big)\, \sigma_3\,.
\ee
%%%%%
In fact the combination
%%%%%
\be
\ft18 (9A_2 -A_1) = \fft{(r-2\ell)(r-\ft12\ell)}{r^2-\ell^2}\, \sigma_3
\ee
%%%%%
is globally defined, and the associated field strength is exact.  The
combination
%%%%%
\be
A= -\ft38 P\, (A_1-A_2) = -\fft{3P\, \ell\, r}{2(r^2-\ell^2)}\, \sigma_3\,,
\label{spincpot}
\ee
%%%%%
which falls off at large $r$ and is singular on the Bolt at $r=2\ell$, 
defines a regular 2-form $F=dA$ whose integrals over the $S^2$ bolt and the
$\R^2$ bundle parameterised by $(r,\psi)$ are given by
%%%%%
\be
\fft1{4\pi}\, \int_{S^2} F = P\,,\qquad \fft1{4\pi}\, \int_{\R^2} F = P\,.
\ee
%%%%%
 
   If there are fermions with charge $e$, the usual Dirac quantisation 
conditions would imply that $2eP$ should be an integer.  However, since
the Taub-BOLT manifold does not admit a spin structure, consistency of the 
fermion wave functions requires that instead, as discussed in 
\cite{hawpop}, we impose the quantisation condition
%%%%%
\be
2eP = q+ \ft12 \,,\label{epquant}
\ee
%%%%%
where $q$ is an integer.  

   A consistency check, analogous to the one described in \cite{hawpop}
for $\CP^2$, can be performed by calculating the Atiyah-Singer index for the
Dirac operator for such charged spinors in the Taub-BOLT manifold.  Thus,
the difference between the numbers of right-handed and left-handed
$L^2$-normalisable zero modes of the charged Dirac operator is given by
%%%%%
\be
n_+-n_- = -\fft1{384\pi^2}\, \int R_{\mu\nu\rho\sigma}\, 
   ^*R^{\mu\nu\rho\sigma}\, \sqrt{g} d^4x + \fft{e^2}{16\pi^2}\, 
\int F_{\mu\nu}\, ^*F^{\mu\nu}\, \sqrt{g} d^4 x - \ft12\eta(0)
\ee
%%%%%
where $-\ft12\eta(0)= -1/12$ is the Atiyah-Patodi-Singer term calculated from
the eigenvalues of the Dirac operator on the boundary.  From (\ref{tbmet})
and the field $F=dA$ coming from (\ref{spincpot}), we find that the Dirac
index is given by
%%%%%
\be
n_+-n_- = 2 e^2 P^2 - \ft18\,.
\ee
%%%%%
Thus we see that with the quantisation condition (\ref{epquant}) 
appropriate to this case where there is no ordinary spin structure, the
index is
%%%%%
\be
n_+-n_-= \ft12 q(q+1)\,,
\ee
%%%%%
which is indeed always an integer.

   One can also check the Hirzebruch index of the operator $d+\delta$,
which gives the difference between the numbers of self-dual and
anti-self-dual harmonic 2-forms.  For the self-dual Taub-NUT
instanton, it is known that this index is $-1$, coming from a $-2/3$ 
contribution from the bulk integral
%%%%%
\be
\fft1{48\pi^2}\,  \int R_{\mu\nu\rho\sigma}\, 
   ^*R^{\mu\nu\rho\sigma}\, \sqrt{g} d^4x \,,\label{hirz}
\ee
%%%%
and a $-1/3$ from the boundary.  The same boundary term arises for 
Taub-BOLT, and thus evaluating the contribution (\ref{hirz}) for
(\ref{tbmet}) we obtain the Hirzebruch signature $1/3-1/3=0$ for the
Taub-BOLT metric.  This is consistent with the existence of the 
one self-dual and one anti-self-dual $L^2$ harmonic forms that we found above.

\end{document}